\documentclass[twocolumn,showpacs,prb,amsmath,amssymb]{revtex4}

\usepackage{graphicx}

\DeclareMathOperator{\Tr}{Tr}
\def\bra#1{\langle#1 |}
\def\ket#1{| #1\rangle}
\newcommand{\braket}[2]{\langle #1 | #2 \rangle}
\def\op#1{\hat{#1}}

\def\diagonalav#1{\langle #1 \rangle_{\mathrm{\scriptscriptstyle D}} }
\def\ggeav#1{\langle #1 \rangle_{\mathrm{\scriptscriptstyle GGE}} }
\def\timeav#1{\langle #1 \rangle_{\mathrm{\scriptscriptstyle time}} }
\newcommand{\ud}{ \mathrm{d} }

\begin{document}

\title{Relaxation and Thermalization after a Quantum Quench:\\ Why Localization is Important}

\author{Simone Ziraldo$^{1,2}$ and Giuseppe E. Santoro$^{1,2,3}$}
\affiliation{
$^1$ SISSA, Via Bonomea 265, I-34136 Trieste, Italy \\
$^2$ CNR-IOM Democritos National Simulation Center, Via Bonomea 265, I-34136 Trieste, Italy \\
$^3$ International Centre for Theoretical Physics (ICTP), P.O.Box 586, I-34014 Trieste, Italy
}
\begin{abstract}
We study the unitary dynamics and the thermalization properties of free-fermion-like Hamiltonians
after a sudden quantum quench, extending the results of S. Ziraldo {\it et al.}
[Phys. Rev. Lett. 109, 247205 (2012)].
With analytical and numerical arguments, we show that the existence of a stationary state and 
its description with a generalized Gibbs ensemble (GGE) depend crucially on the observable considered (local versus extensive)
and on the localization properties of the final Hamiltonian.
We present results on two one-dimensional (1D) models, the disordered 1D fermionic chain with long-range hopping and 
the disordered Ising/$XY$ spin chain.
We analytically prove that, while time averages of one-body operators are perfectly reproduced by GGE (even for finite-size systems, if time integrals
are extended beyond revivals), time averages of many-body operators might show clear deviations from the GGE prediction when disorder-induced
localization of the eigenstates is at play.
\end{abstract}


\date{\today}
\maketitle

\section{Introduction} \label{sec:intro}

The concept of ergodicity is at the core of classical statistical mechanics: it establishes a connection 
between long-time averages of observables and statistical ensemble averages \cite{Huang:book}. 
The extension of the ergodic theorem to quantum mechanics was pioneered by 
von Neumann~\cite{vonNeumannHTheorem,vonNeumannTranslation} in a seminal paper
on the unitary dynamics of closed quantum systems. 
The experimental possibility of studying the nonequilibrium dynamics of ``virtually'' isolated quantum systems  
-- most notably cold atomic species in optical lattices~\cite{Bloch_RMP08,Lewenstein_AP07} -- has stimulated
new interest in this issue. 
The simplest setting for such a nonequilibrium situation is that of a sudden quench of the Hamiltonian, $\op{H}_0 \to \op{H}$, 
governing the dynamics of the isolated quantum system: 
the after-quench ($t>0$) unitary evolution is simply $|\Psi(t)\rangle=e^{-i\op{H}t}|\Psi_0\rangle$ but the initial state 
$|\Psi_0\rangle$ --- for instance, the ground state of $\op{H}_0$ --- is generally a complicated superposition of the eigenstates of $\op{H}$. 
Since the energy is conserved, $\langle \Psi(t) | \op{H} |\Psi(t)\rangle = \langle \Psi_0 | \op{H} |\Psi_0\rangle$, it is reasonable
to expect that a ``generic'' ergodic evolution in the Hilbert space will lead to time averages which are reproduced by the microcanonical ensemble:
this is what von Neumann discussed for ``macroscopic'' observables \cite{vonNeumannHTheorem,vonNeumannTranslation},  
and is generally expected to occur~\cite{Deutsch_PRA91,Sred_PRE94,Rigol_Nat}, independently of the initial state $|\Psi_0\rangle$. 

Classical physics teaches us also when to expect violations of ergodicity: on one extreme, for systems that are integrable \cite{Saletan:book} 
or close enough to being integrable \cite{Fermi-Pasta-Ulam,KAM-papers}; on the other, for systems with a glassy dynamics, be it due to interactions
providing dynamical constraints \cite{Biroli_PRL02}, or to genuine disorder \cite{Parisi_Mezard_Virasoro:book}.
Quantum mechanically, dangers to ergodicity come from very similar sources: integrability, interactions, and disorder.
Integrability implies the existence of many constants of motion $\op{I}_{\mu}$, and this clearly restricts the ergodic exploration of the microcanonical
energy shell, leading to what one might call a ``breakdown of thermalization'' \cite{Rigol_PRL09, Rigol_PRA09}. 
It often results in a kind of ``generalized thermalization'' described by a statistical ensemble which maximizes entropy in the presence of
the constraints $\langle \op{I}_{\mu}\rangle$, an ensemble introduced long ago by Jaynes \cite{Jaynes_PR57} and
known as generalized Gibbs ensemble (GGE) ~\cite{Rigol_PRB06,Rigol_PRL07,Barthel2010,Calabrese2011,Cazalilla2012}
\begin{equation}
\op{\rho}_{\mathrm{\scriptscriptstyle GGE}} \equiv \frac{e^{-\sum_\mu \lambda_\mu \op{I}_\mu}}{ \Tr \left[ e^{-\sum_\mu \lambda_\mu \op{I}_\mu} \right]} \;,
\end{equation}
where $\lambda_\mu$ are Lagrange multipliers which constrain the mean value of each $\op{I}_\mu$ to its $t=0$ value: 
$\bra{\Psi_0} \op{I}_\mu \ket{\Psi_0} = \Tr\left[ \op{\rho}_{\mathrm{\scriptscriptstyle GGE}} \op{I}_\mu \right]$.

On the theory side, the approach to equilibrium has been carefully investigated for 
one-dimensional Bose-Hubbard models describing quench experiments in optical 
lattices and superlattices~\cite{Cramer2008a,Cramer2008,Cramer2010,Flesch2008}. 
Starting from nonhomogeneous initial states (density waves), such as $|\Psi_0\rangle=|1,0,1,0,\dots,1,0\rangle$, 
and evolving the system with a Bose-Hubbard $\op{H}=-J\sum_i(\op{b}_{i+1}^\dagger\op{b}_i + H.c.) + (U/2) \sum_i \op{n}_i(\op{n}_i-1)$
both at the integrable points  $U=0$ (free bosons) and $U=\infty$ (free hard-core bosons/spinless fermions), 
where analytic solutions are possible \cite{Flesch2008}, and at general (nonintegrable) values of $U$ 
(through time-dependent density-matrix renormalization group~\cite{Daley_JSTAT04, White_PRL04}), 
the physical picture emerging has led to the so-called {\em local relaxation conjecture}\cite{Cramer2008a}: 
although the system is, {\em sensu stricto}, in a pure state $\op{\rho}(t) = |\Psi(t)\rangle\langle \Psi(t)|$, when measured upon locally in a finite region $S$, 
the resulting (mixed) reduced density matrix $\op{\rho}_{S}(t)={\rm Tr}_{L\setminus S} \op{\rho}(t)$ relaxes towards a stationary Gibbs state
of maximum entropy compatible with all the constants of motion.
This relaxation is strongly tight to the ``locality'' of the observable $\op{O}$ one measures, and results from ``information transfer'' carried by the excitations
along the system \cite{Cramer2008a,Cramer2008,Flesch2008}, which eventually ``thermalizes'' any finite region $S$, the rest of the system acting as an effective bath for $S$.
Remarkably, such an approach to equilibrium does not require 
time averages~\cite{Cramer2008a}.
Experimentally, a fast dynamical relaxation was recently observed~\cite{Trotzky2012} in a system of cold atoms, where the long-time stationary state was compatible with GGE.

A far more dangerous type of ``ergodicity crisis'' derives, apparently, from dynamical constraints imposed by interactions\cite{Carleo_SRep12}: 
when quenching a Bose-Hubbard model starting, for instance, from $|\Psi_0\rangle=|2,0,2,0,\dots,2,0\rangle$ 
--- i.e, with an average density of bosons per site $n=1$ where, in equilibrium, a Mott phase transition occurs for $U>U_c\sim 3.5 J$ --- 
the ensuing dynamics leads to a fast relaxation/thermalization for quenches at small $U$, while the relaxation is extremely slow
(and the more so, the more the size of the system increases) and the dynamics appears effectively {\em freezed} for $U>U^{\mathrm{dyn}}_{c}$ \cite{Carleo_SRep12}.
Concerning ergodicity breaking due to genuine disorder, both Anderson localization, at the single-particle level~\cite{Anderson_PR58}, 
and many-body localization, in the presence of interactions~\cite{Basko_AP06}, are 
well-known examples of disorder-induced phenomena occurring in equilibrium physics.
Quantum quenches in the presence of disorder and interactions have also been studied, in the framework of many-body localization \cite{Oganesyan_PRB07, Pal_PRB10,Canovi_PRB11,Canovi_NJP12}, 
but the physical picture is far from being fully understood.

While breaking of translational invariance and disorder in the initial state $|\Psi_0\rangle$ have apparently 
little or no effect on the ensuing relaxation towards a stationary state if the after-quench Hamiltonian $\op{H}$ 
is translationally invariant~\cite{Gangardt2008,Caneva_JSM11,Calabrese2011,Cazalilla2012,Ziraldo_PRL12},
disorder in the final Hamiltonian $\op{H}$ seems to play a quite crucial role: this has been seen in numerical
studies of nonintegrable fermion models with disordered long-range hopping and nearest-neighbor interaction~\cite{Khatami},
and has been recently corroborated through analytical arguments on integrable quantum spin chains of the Ising/$XY$ class~\cite{Ziraldo_PRL12}.

Our present study extends that of Ref.~\onlinecite{Ziraldo_PRL12}, where it was shown that, in the presence of disorder in $\op{H}$,
the long-time after-quench dynamics {\em does not relax} towards a stationary state, and time fluctuations generally persist in
the expectation values of local operators, even in the thermodynamic limit; 
this is, essentially, due to the presence of a {\em pure-point spectrum} of $\op{H}$ 
associated to localized wave functions, as opposed to the smooth continuum of a system with extended states.
One can view this persistence of time fluctuations in local measurements as a result of the inability of the system to carry information around \cite{Cramer2008a},
due to localization.
Time averages are, therefore, mandatory in comparing dynamical quantities to statistical ensemble averages.
But here a further surprise emerges, which will be the main focus of this study: 
while time averages of one-body local observables 
are perfectly well reproduced by the relevant statistical ensemble --- the GGE, being the considered models, essentially, free-fermion ones ---
time averages of many-body local operators show clear deviations from the corresponding GGE prediction, in agreement with recent numerical findings~\cite{Gramsch}.   

We will exemplify these ideas on a model of disordered one-dimensional fermions with long-range hoppings, 
showing a transition between extended and localized eigenstates, and on disordered Ising/$XY$ models. 
We will start, Sec.~\ref{sec:problem}, by stating in a more precise way the problem we want to analyze. 
Next, in Sec.~\ref{sec:models}, we describe the models we have investigated. In Sec.~\ref{sec:GGE-one-body} we will analyze the essential reason why GGE
works perfectly well in predicting infinite-time averages of {\em one-body operators}, even for finite systems (as long as one integrates over revivals).
In Sec.~\ref{sec:GGE-many-body} we will discuss why this is not generally the case for many-body operators, unless time fluctuations of one-body Green's functions
vanish for large $t$.
Next, in Secs.~\ref{sec:spinless} and \ref{sec:Ising} we will present our results for the two models we have considered.
Section~\ref{discussion:sec} contains a final discussion, and our conclusions and perspectives.

\section{Statement of the problem} \label{sec:problem}
Let us start by defining the problem. Consider a standard quantum quench:  an initial state
$|\Psi_0\rangle$, ground state of some Hamiltonian $\op{H}_0$, evolves under a different
time-independent Hamiltonian $\op{H}$.
Given an observable $\hat{O}$, its quantum average can be separated in two terms 
(we take $\hbar=1$):
\begin{equation} \label{defO:eqn}
O(t) \equiv \langle \Psi_0 | e^{i\op{H}t} \hat{O} e^{-i\op{H}t} | \Psi_0\rangle = \timeav{\op{O}} + \delta O(t)  \;,
\end{equation}
where the time-independent part $\timeav{\op{O}}$ results from an infinite-time average of $O(t)$ 
\begin{equation}
\timeav{ \op{O} } = \lim_{T\to \infty} \frac{1}{T} \int_0^T \! dt \; O(t) \;,
\end{equation}
while $\delta O(t)$ represents a fluctuating part (with a vanishing time average).
An alternative standard decomposition of $O(t)$ proceeds by introducing the (many-body) eigenstates
$|\alpha\rangle$ of $\op{H}$, with energy $E_{\alpha}$, and inserting resolutions of the identity in Eq.~\eqref{defO:eqn}.
Defining $C_{\alpha}\equiv \langle \alpha | \Psi_0\rangle$ and $O_{\alpha'\alpha} \equiv \bra{\alpha'} \hat{O} \ket{\alpha}$, one gets
\begin{eqnarray}
O(t) &=& \sum_{\alpha}
       |C_{\alpha}|^2 O_{\alpha\alpha} +
        \sum_{\alpha'\ne \alpha} e^{i(E_{\alpha'}-E_{\alpha})t}
       C_{\alpha'}^* O_{\alpha'\alpha} C_{\alpha}  \nonumber \\
     &=& \diagonalav{\hat{O}} +   \int_{-\infty}^{+\infty} \! \! \! d\Omega \, e^{-i\Omega t} F_O(\Omega) \, , \nonumber
\end{eqnarray}
where the first term, $\diagonalav{\hat{O}}$, is the so-called diagonal average \cite{vonNeumannHTheorem,vonNeumannTranslation,Rigol_Nat},
while the second, time-dependent, part has been recast as a Fourier transform of a weighted joint (many-body) density of states
$F_O(\Omega) \equiv \sum_{\alpha'\ne\alpha} C_{\alpha'}^* O_{\alpha'\alpha} C_{\alpha} \delta\left(\Omega-E_{\alpha}+E_{\alpha'}\right)$.
With the quite usual assumption of no energy degeneracy, $E_{\alpha'}\neq E_{\alpha}$ if $\alpha'\neq\alpha$, one can quickly prove that 
the diagonal average indeed coincides with the long-time average \cite{vonNeumannHTheorem,vonNeumannTranslation,Rigol_Nat}:
\begin{equation}
 \diagonalav{\hat{O}} = \timeav{ \op{O} } = \lim_{T\to \infty} \frac{1}{T} \int_0^T \! dt \; O(t) \;,
\end{equation}
while time fluctuations are given by the Fourier transform of $F_O(\Omega)$:
\begin{equation}
\delta O(t) = \int_{-\infty}^{+\infty} \! \! \! d\Omega \, e^{-i\Omega t} F_O(\Omega) \;.
\end{equation}
As discussed in Ref.~\onlinecite{Ziraldo_PRL12}, the behavior of the fluctuating 
part $\delta O(t)$, relaxing and decaying to 0 
or remaining finite (with persistent oscillations) for $t\to \infty$, is strongly tied to 
the ``smoothness'' of $F_O(\Omega)$ in the
thermodynamic limit (for finite systems, $F_O(\Omega)$ is always a series of 
discrete Dirac $\delta$'s, hence
$\delta O(t)$ will never go to zero for $t\to \infty$, and revivals will appear).
Indeed,  $\delta O(t)$ will decay to zero for large $t$ if $F_O(\Omega)$ is smooth enough, due to the destructive interference induced in 
the $\Omega$ integral by the strongly oscillating phase $e^{-i\Omega t}$ (Riemann-Lebesgue lemma);
on the contrary, disorder and an important {\em pure-point spectrum} part, i.e., $\delta$ functions associated to localized eigenstates which do
not merge smoothly into a continuum, will lead to persistent time fluctuations $\delta O(t)$ for local operators \cite{Ziraldo_PRL12}.

In the rest of the paper we will concentrate on two ``solvable'' models with disorder which will be presented in Sec.~\ref{sec:models}: 
disordered one-dimensional fermions with long-range hoppings (which can show either 
power-law localized states or extended ones,
depending on a parameter controlling the long-range hopping variance), and disordered Ising/$XY$ model. 
Both, being quadratic fermionic models, can be quite effectively numerically diagonalized, i.e., one can find
the one-body spectrum $\epsilon_\mu$ and the corresponding quasiparticle creation operator $\op{\gamma}_{\mu}^\dagger$
to express $\op{H} = \sum_{\mu} \epsilon_\mu \op{\gamma}_{\mu}^\dagger \op{\gamma}_{\mu}$.
This, in turn, allows us to calculate the fermionic local one-body Green's functions:
\begin{eqnarray}
G_{j_1j_2}(t) &\equiv& \bra{\Psi(t)} \op{c}^\dagger_{j_1} \op{c}_{j_2} \ket{\Psi(t)} \nonumber \\
F_{j_1j_2}(t)  &\equiv& \bra{\Psi(t)} \op{c}^\dagger_{j_1} \op{c}_{j_2}^\dagger \ket{\Psi(t)} \;, 
\end{eqnarray}
where $c^{\dagger}_j$ creates a fermion at site $j$.
We will show that, in view of Wick's theorem, a key issue in understanding the validity of GGE for general many-body observables in such free-fermion-like Hamiltonians 
has to do with the nature of the long-time fluctuations of the one-body Green's functions.
That might seem a simple matter to explore, but unfortunately, in the general disordered case, the one-body Green's functions are essentially impossible 
to obtain analytically, and their numerical study is often elusive: you can only study a finite system-size $L$ for a finite time $t$, and
whether the fluctuations will eventually vanish or not for $L\to \infty$ (first) and $t\to \infty$ (after, otherwise you always
get revivals) is often hard to tell.
%
%
It turns out that a useful tool to distinguish the presence or absence of time fluctuations for large $t$ 
is given by the time-averaged fluctuations:
\begin{equation}
\delta^2_{O} \equiv \lim_{T \rightarrow \infty} \frac{1}{T} \int_0^T
\ud t
 \left| \delta O(t) \right|^2 \;,
\end{equation}
which is zero when the fluctuations vanish for large $t$, and is finite otherwise. 
We will show how to make progress analytically, for this quantity, to pin point the behavior of the large-$t$ fluctuations of one-body Green's functions. 
This analytical progress will require an assumption of {\em absence of gap degeneracies}, 
i.e., $\epsilon_{\mu_1} - \epsilon_{\mu_2} = \epsilon_{\mu_3} - \epsilon_{\mu_4}$ only when $\mu_1=\mu_2$ and $\mu_3=\mu_4$ or $\mu_1=\mu_3$ and $\mu_2=\mu_4$,
restricted, however, to the {\em one-body spectrum}, an assumption that can be argued to be a reasonable one, especially in the presence of disorder. 
Its general validity for many-body energy eigenvalues, on the contrary, is rather tricky
(see discussion in Sec.~\ref{discussion:sec}).

In order to get smoother results we will perform averages of $\delta^2_{O}$ over different 
realizations of disorder. 
We stress, however, that the disorder average will
always be performed {\em after} the computation of $\delta^2_{O}$ for each realization: 
performing the disorder average before taking the squared time integral, i.e., on $\delta O(t)$, would effectively kill the time fluctuations
$\delta O(t)$, by a kind of ``self-averaging'' \cite{Binder_Young:review}.

\section{Models}\label{sec:models}
%
In this work we concentrate on two ``solvable'' models possessing a simple fermionic description.
The first model describes spinless fermions hopping on a chain~\cite{Mirlin}: 
\begin{equation}\label{eq:fermham}
\op{H}_{\mathrm{hop}} = \sum_{j_1j_2} J_{j_1 j_2} (\op{c}_{j_1}^\dagger \op{c}_{j_2} + \mathrm{H.c.}) \;,
\end{equation}
where $\op{c}_{j}^\dagger$ ($\op{c}_{j}$) creates (destroys) a fermion at site $j$ and $J_{j_1j_2}$ is a (real) hopping integral
between sites $j_1$ and $j_2$. We will in general take the $J_{j_1 j_2}$'s to be random and long ranged, with a 
Gaussian distribution of zero mean, $\langle J_{j_1j_2} \rangle = 0$, and variance given by:
\begin{equation}
\langle J_{j_1j_2}^2 \rangle = \frac{1}{1+\left( \frac{|j_1-j_2|}{\beta} \right)^{2\alpha}} \,.
\end{equation}
Here $\alpha$ is a real positive parameter setting how fast the hoppings' variance decays with distance.
The peculiarity of this long-range-hopping model is that, regardless of the value of $\beta$ (which hereafter is fixed to $1$),
it has an Anderson transition from (metallic) extended eigenstates, for $\alpha<1$, to (insulating) power-law localized eigenstates for 
$\alpha>1$ \cite{Mirlin,Cuevas,Varga}.
Physically, this is due to the fact that, for small $\alpha$, long-range hoppings are capable of overcoming the localization 
due to disorder. 
The clean nearest-neighbor hopping model is recovered by taking 
$J_{j_1j_2} = -\delta_{j_2,j_1\pm 1}$, where $\delta_{i,j}$ is the Kronecker delta: 
we will always use this choice for the initial Hamiltonian $\op{H}_0$, with the corresponding ground state $|\Psi_0\rangle$ 
being the filled Fermi sea. 
(The reason behind this simple choice for $\op{H}_0$ is that the long-time fluctuation properties do not depend, qualitatively,
on the initial Hamiltonian being ordered or not, see Ref.~\onlinecite{Ziraldo_PRL12}). 
Being quadratic in the fermion operators, $\op{H}_{\mathrm{hop}}$ can be diagonalized for any chain of size $L$
in terms of new fermionic operators 
\begin{equation} \label{cmu-cj:eqn}
 \op{c}_\mu^\dagger =  \sum_{j=1}^L u_{j\mu} \op{c}_j^\dagger \;,
\end{equation}
where $u_{j\mu}$ are the wave functions of the eigenmodes of energy $\epsilon_\mu$:
$\op{H}_{\mathrm{hop}} = \sum_{\mu} \epsilon_\mu \op{c}_\mu^\dagger \op{c}_\mu $. 
The energies $\epsilon_\mu$ and the associated wave functions  $u_{j\mu}$ are obtained, for any
given realization of the hoppings $J_{j_1j_2}$ in a chain of size $L$ with open boundary
conditions, by numerically diagonalizing the  $L \times L$ one-body hopping matrix.

The second Hamiltonian we considered describes a disordered Ising/$XY$ chain in a transverse field \cite{Fisher_PRB95}:
\begin{equation} \label{eq:isingham}
\op{H}_{XY} = - \sum_{j=1}^{L} \left( J_j^x \op{\sigma}^x_j\op{\sigma}^x_{j+1} + J_j^y \op{\sigma}^y_j \op{\sigma}^y_{j+1} \right) - 
                            \sum_{j=1}^L h_j \op{\sigma}_i^z \,,
\end{equation}
where $L$ is the size of the chain, $\op{\sigma}_j^\mu$ ($\mu = x,y, z$) are spin-$1/2$ Pauli matrices 
for the $j$ site, and periodic boundary conditions are assumed, $\op{\sigma}_{L+1}^\mu = \op{\sigma}_{1}^\mu$.
$J_j^x$, $J_j^y$ and $h_j$ are real and describe, respectively, the nearest-neighbor spin couplings and the transverse magnetic field.
A quadratic fermionic Hamiltonian is obtained here by applying a Jordan-Wigner transformation \cite{Lieb_AP61}
$ \op{c}_l \equiv \op{\sigma}_l^- \exp \left( i \pi \sum_{j=1}^{l-1} \op{\sigma}_j^+ \op{\sigma}_j^- \right)$,
where $ \op{\sigma}_j^\pm \equiv (\op{\sigma}_j^x \pm i \op{\sigma}_j^y)/2$:
\begin{eqnarray}
\op{H}_{XY} = &-& \sum_{j=1}^{L} J_j \left( \op{c}_j^\dagger \op{c}_{j+1} + \gamma
                                                                           \op{c}_j^\dagger \op{c}_{j+1}^\dagger + \mathrm{H.c}. \right) \nonumber \\
                        &-& \sum_{j=1}^L h_j (2 \op{c}^\dagger_j \op{c}_j - 1)    \,,
\end{eqnarray}
where $J_j = J_j^x + J_j^y$, and the anisotropy parameter $\gamma$ is such that $J_j^x =J_j(1+\gamma)/2$ 
and $J_j^y =J_j(1-\gamma)/2$. In the relevant sub-sector with an even number of fermions, we have to
apply anti-periodic boundary conditions $\hat{c}_{L+1} = - \hat{c}_{1}$ \cite{Lieb_AP61}. 
At variance with the fermion-hopping case, Eq.~\eqref{eq:fermham}, there are now, for $\gamma\ne 0$,
BCS terms $\op{c}_j^\dagger \op{c}_{j+1}^\dagger$ which create (and destroy) pairs of fermions. 
Using a Bogoliubov rotation we therefore define new fermions \cite{Young1996,Young1997}
\begin{equation}  \label{eq:defQuasiPart}
\op{\gamma}_\mu^\dagger = \sum_{j=1}^L \left( u_{j\mu} \op{c}_j^\dagger + v_{j \mu} \op{c}_j \right)  \;,
\end{equation}
which diagonalize the Hamiltonian 
$\op{H}_{XY} = \sum_{\mu=1}^{L} \epsilon_\mu \left( \op{\gamma}^\dagger_\mu \op{\gamma}_\mu - 1/2 \right)$. 
The (positive) eigenvalues $\epsilon_\mu/2$ and the associated eigenfunctions $(u_{j\mu}, v_{j\mu})$
are obtained, once again, by diagonalizing a $2L\times 2L$ one-body
matrix~\cite{Young1996,Young1997,Caneva_PRB07}. 
Notice also the strict particle-hole symmetry present even in the general disordered 
case~\cite{Young1996,Young1997}:
for every positive eigenvalue $\epsilon_\mu/2>0$, with associated $(u_{j\mu}, v_{j\mu})$, there is a negative eigenvalue 
$-\epsilon_\mu/2$ associated to $(v_{j\mu}^\ast, u_{j\mu}^\ast)$.

When considering quenches for this Hamiltonian, we always start from a clean $\op{H}_0$, with 
$J_j=1$, $\gamma=1$ (Ising case), and $h_j=h_0$, while, for the final disordered Hamiltonian, we take $J_j = 1 + \epsilon \eta_j$, $\gamma=1$, and $h_j = h + \epsilon \xi_j$, where $\epsilon$ sets the disorder strength
and $\eta_j$, $\xi_j$ are uncorrelated uniform random numbers in $[-1, 1[$. 
The aim of our analysis is to study the behavior of different disorder realizations for very large $L$ (ideally, in the thermodynamic limit). 
Practically, the largest size we will consider is $L=2048$. Given a disorder realization for $L=2048$, we will
generate corresponding realizations at smaller $L$ by cutting away the same amount of sites from the two edges.
In this way, we obtain a smoother behavior for all quantities versus $L$.
For every quantity $x$ considered, we have checked its probability distribution $P(x)$ for different realizations of disorder.
Sometimes $P(x)$ deviates strongly from a Gaussian distribution, and is nearly (although not precisely) lognormal, i.e.,
it is $\log(x)$ which is approximately Gaussian distributed. 
In such a situation, we will calculate and plot the median (i.e., the geometric mean) 
$\left[ x \right]_{\mathrm{av}}^\mathrm{G} = \exp ( \left[ \log x \right]_{\mathrm{av}} )$ 
and the geometric standard deviation $\exp ( \sigma[\log x] )$,
rather than the usual (arithmetic) mean $\left[ x \right]_{\mathrm{av}}$, and its standard deviation $\sigma[x]$ 
(in the plots, the error bars will then go from 
$\left[ x \right]_{\mathrm{av}}^\mathrm{G} \exp ( -\sigma[\log x] ) $
to  $\left[ x \right]_{\mathrm{av}}^\mathrm{G} \exp ( \sigma[\log x] )$).

For both the models considered above, after diagonalization, the Hamiltonian is
expressed as:
\begin{equation}\label{eq:ham} 
\op{H} = \sum_{\mu} \epsilon_\mu \op{\gamma}_{\mu}^\dagger \op{\gamma}_{\mu} + E_0 \, ,
\end{equation}
where $\epsilon_\mu$ is the positive excitation energy of the state $\op{\gamma}_\mu^{\dagger}|0\rangle$
and $E_0$ is the energy of the state $|0\rangle$ annihilated by all the $\op{\gamma}_{\mu}$.
Notice that, in diagonalizing $\op{H}_{\mathrm{hop}}$, Eq.~\eqref{eq:fermham}, one generally
obtains some negative $\epsilon_\mu$: in such a case, it is enough to perform a
particle-hole transformation $\gamma_{\mu} = \op{c}_{\mu}^\dagger$ to change the sign 
of $\epsilon_\mu$. Physically, that implies that all negative energies are occupied in the
ground state $|0\rangle$, and the resulting excitations describe particles or holes.

In the following two sections we will use expression \eqref{eq:ham} for $\op{H}$:
all the observations made are valid for both the models we have just described.
Notice that the number operators $\op{\gamma}_\mu^\dagger\op{\gamma}_\mu$ 
commute with $\op{H}$ in Eq.~\eqref{eq:ham} and are therefore obvious constants of 
motion in the GGE averages, $\op{I}_\mu=\op{\gamma}_\mu^\dagger\op{\gamma}_\mu$. 

\section{Why GGE works for one-body observables} \label{sec:GGE-one-body}
%
In this section we are going to show that, for a free fermion Hamiltonian of the form
given in Eq.~\eqref{eq:ham}, the long-time average (and the diagonal average) of any one-body operator coincides 
with the corresponding GGE average, for any system size $L$, and any possible quench:
\begin{equation} \label{1-body-averages:eqn}
    \timeav{\op{O}_{\mathrm{1-body}}} =  \diagonalav{\op{O}_{\mathrm{1-body}} } = \ggeav{\op{O}_{\mathrm{1-body}} } \;, 
\end{equation}
with some important remarks to be made when there are degeneracies in the one-body spectrum (see below).
As we shall see, this can be traced back to the constraints that the GGE sets through the constants of motion $\op{I}_\mu$.
We stress that, remarkably, this equality holds even for a finite-size chain $L$,
while, usually, statistical ensembles need a thermodynamic limit.
Examples of one-body operators are, in real space, $\op{c}^\dagger_{j_1} \op{c}_{j_2}$ or 
$\op{c}^\dagger_{j_1} \op{c}^\dagger_{j_2}$, the local density $\op{n}_j \equiv \op{c}^\dagger_j \op{c}_j$, 
the density $ \op{n} \equiv \sum_j \op{n}_j/L$, and, in momentum space,
$\op{c}^\dagger_{k} \op{c}_{k}$ (where $\op{c}^{\dagger}_k =  \sum_j e^{ikj} \op{c}^{\dagger}_j/\sqrt{L}$), etc.
More generally, a one-body fermionic operator can always be written, neglecting irrelevant constants
and rewriting the $\op{c}_j$'s in terms of the $\op{\gamma}_\mu$ (inverting Eqs.~\eqref{cmu-cj:eqn} or \eqref{eq:defQuasiPart}), as:
\begin{align}
\op{O}_{\mathrm{1-body}} = &  \sum_{\mu_1\mu_2} A_{\mu_1 \mu_2}  
	\op{\gamma}_{\mu_1}^\dagger \op{\gamma}_{\mu_2}
+ \notag \\
& +
\sum_{\mu_1\mu_2} B_{\mu_1 \mu_2}  
	\op{\gamma}_{\mu_1}^\dagger \op{\gamma}_{\mu_2}^\dagger
+
\sum_{\mu_1\mu_2} D_{\mu_1 \mu_2}  
	\op{\gamma}_{\mu_1} \op{\gamma}_{\mu_2} \;,
\label{eq:one-body} 
\end{align}
where $A$, $B$, and $D$ are $L \times L$ matrices.
Let us start showing that  $\diagonalav{ \op{O}_{\mathrm{1-body}} } = 
\ggeav{ \op{O}_{\mathrm{1-body}} }$.
If $\ket{\alpha}=\op{\gamma}_{\mu_1}^\dagger \op{\gamma}_{\mu_2}^\dagger  \cdots |0\rangle$ denotes a general many-body eigenstate of $\op{H}$, 
then clearly only the diagonal elements of $A$ enter in the diagonal matrix element:
\begin{equation}
   \bra{\alpha}  \op{O}_{\mathrm{1-body}} \ket{\alpha} = \sum_\mu A_{\mu\mu} \bra{\alpha}  \op{\gamma}_{\mu}^\dagger \op{\gamma}_{\mu} \ket{\alpha} \;,
\end{equation}
where $\bra{\alpha}  \op{\gamma}_{\mu}^\dagger \op{\gamma}_{\mu} \ket{\alpha} =  0, 1$ is the occupation number of 
eigenmode $\mu$ in the eigenstate $\ket{\alpha}$.
In terms of $\bra{\alpha}  \op{O}_{\mathrm{1-body}} \ket{\alpha}$, the diagonal average of  $\op{O}_{\mathrm{1-body}}$ is readily expressed as:
\begin{align}
\diagonalav{\op{O}_{\mathrm{1-body}}}  & = \sum_{\alpha} \sum_{\mu} 
|C_\alpha|^2 A_{\mu \mu} \bra{\alpha}  \op{\gamma}_{\mu}^\dagger \op{\gamma}_{\mu} \ket{\alpha} \notag \\
& =\sum_{\mu} A_{\mu \mu} \sum_{\alpha} 
 \braket{\Psi_0}{\alpha} \bra{\alpha} \op{\gamma}_{\mu}^\dagger \op{\gamma}_{\mu} \ket{\alpha} \braket{\alpha}{\Psi_0} \notag \\ 
 & =\sum_{\mu} A_{\mu \mu} \sum_{\alpha\alpha'} 
 \braket{\Psi_0}{\alpha} \bra{\alpha} \op{\gamma}_{\mu}^\dagger \op{\gamma}_{\mu} \ket{\alpha'} \braket{\alpha'}{\Psi_0} \notag \\ 
& = \sum_{\mu} A_{\mu \mu} \bra{\Psi_0} \op{\gamma}_{\mu}^\dagger \op{\gamma}_{\mu} \ket{\Psi_0} \;,
\end{align}
where we have added and extra sum over $\alpha'$, using 
$\bra{\alpha} \op{\gamma}_{\mu}^\dagger \op{\gamma}_{\mu} \ket{\alpha'}=\delta_{\alpha,\alpha'} \bra{\alpha} \op{\gamma}_{\mu}^\dagger \op{\gamma}_{\mu} \ket{\alpha}$,
and then recognized two resolutions of the identity $\sum_{\alpha} \ket{\alpha}\bra{\alpha}$.
Notice, therefore, that the initial state enters only through $\bra{\Psi_0} \op{\gamma}_{\mu}^\dagger \op{\gamma}_{\mu} \ket{\Psi_0}$, 
i.e., exactly the constants of motion which are constrained and reproduced by the GGE averages:
$\ggeav{\op{\gamma}_\mu^\dagger \op{\gamma}_\mu}=\bra{\Psi_0} \op{\gamma}_\mu^\dagger \op{\gamma}_\mu \ket{\Psi_0}$.
The conclusion is therefore simple, as the GGE average of $\op{O}_{\mathrm{1-body}}$ is:
\begin{align}
\ggeav{\op{O}_{\mathrm{1-body}}} & = 
\sum_\mu A_{\mu\mu} \ggeav{\op{\gamma}_\mu^\dagger \op{\gamma}_\mu} \nonumber \\ &= 
\sum_\mu A_{\mu\mu} \bra{\Psi_0} \op{\gamma}_\mu^\dagger 
\op{\gamma}_\mu \ket{\Psi_0} \;,
\label{eq:gge1body}
\end{align}
where we used that, by construction of the GGE, 
$\ggeav{\op{\gamma}_{\mu_1}^\dagger \op{\gamma}_{\mu_2\ne \mu_1}}=
\ggeav{\op{\gamma}_{\mu_1}^\dagger \op{\gamma}_{\mu_2}^\dagger} =\ggeav{\op{\gamma}_{\mu_1} \op{\gamma}_{\mu_2}} =0$.

Concerning the equality $\timeav{\op{O}_{\mathrm{1-body}}} =  
\diagonalav{ \op{O}_{\mathrm{1-body}} }$, we should pay attention to the cases in which $\op{H}$ has degenerate single-particle eigenvalues, 
$\epsilon_{\mu_1} = \epsilon_{\mu_2\neq \mu_1}$ (for instance, when $\op{H}$ is disorder-free). In these cases, the time average of ${O}_{\mathrm{1-body}}(t)$ suppresses all the oscillatory factors
$e^{\pm i(\epsilon_{\mu_2}+\epsilon_{\mu_1})t}$ occurring in the $B$ and $D$ terms of Eq.~\eqref{eq:one-body}, but all the factors $e^{i(\epsilon_{\mu_2}-\epsilon_{\mu_1})t}$ corresponding
to degenerate eigenvalues appearing in the $A$ terms survive. Therefore:
\begin{eqnarray}
&& \timeav{\op{O}_{\mathrm{1-body}}} =  \sum_{ (\mu_1,\mu_2)}^{(\epsilon_{\mu_1} = \epsilon_{\mu_2} )} A_{\mu_1 \mu_2} \bra{\Psi_0} \op{\gamma}_{\mu_1}^\dagger \op{\gamma}_{\mu_2} \ket{\Psi_0} 
\hspace{12mm} \nonumber \\
&& =\diagonalav{ \op{O}_{\mathrm{1-body}} } 
+ \sum_{ (\mu_1,\mu_2\neq \mu_1) }^{(\epsilon_{\mu_1} = \epsilon_{\mu_2} )} A_{\mu_1 \mu_2} \bra{\Psi_0} \op{\gamma}_{\mu_1}^\dagger \op{\gamma}_{\mu_2} \ket{\Psi_0} \;,
\label{eq:timeaverage1body}
\end{eqnarray}
where we have singled out the diagonal elements, and the second sum runs over all the degenerate pairs $(\mu_1,\mu_2\neq \mu_1)$ such that
$\epsilon_{\mu_1} = \epsilon_{\mu_2}$.
Because of degeneracies, however, there is more freedom in the choice of the fermionic operators $\op{\gamma }^\dagger_{\mu}$:
we can always perform a unitary rotation in each degenerate subspace in such a way that 
$\bra{\Psi_0} \op{\gamma }^\dagger_{\mu_1} \op{ \gamma }_{\mu_2} \ket{\Psi_0}=0$ for $\mu_1 \neq \mu_2$.
With such a choice of the $\op{\gamma }^\dagger_{\mu}$'s, the extra terms in Eq.~\eqref{eq:timeaverage1body} 
vanish, and we recover the initial statement in Eq.~\eqref{1-body-averages:eqn}, i.e., for any size and any quench,
the long-time average of any one-body operator is equal to the GGE one.
We stress the fact that, for any finite system, $O_{\mathrm{1-body}}(t)$ will have recurrent fluctuations $\delta O_{\mathrm{1-body}}(t)$, the so-called {\em returns} or {\em revivals},
due to the discreteness of the finite-size spectrum: nevertheless, integrating over all times (across revivals) is guaranteed to reproduce the
GGE average:
\[ \lim_{T\to \infty} \frac{1}{T} \int_0^T \! dt \; O_{\mathrm{1-body}}(t) = \ggeav{\op{O}_{\mathrm{1-body}}} \;. \]
This statement, however, does not imply that the fluctuating part $\delta O_{\mathrm{1-body}}(t)$ decreases to $0$ for $t\to \infty$,
as indeed evident from the presence of finite-size revivals. As we shall see, $\delta O_{\mathrm{1-body}}(t)$ might indeed persist for
all times even in the thermodynamic limit, when $\op{H}$ is disordered: this in turn implies that $\lim_{t\to\infty}  O_{\mathrm{1-body}}(t)$
might not exist in some cases, preventing a straightforward application of Wick's theorem to extend the equalities of averages
in Eq.~\eqref{1-body-averages:eqn} to many-body operators.

\section{GGE for many-body observables} \label{sec:GGE-many-body}
%
In the previous section we have shown that the GGE average of a one-body operator coincides exactly with its long-time average.
Here we will see that, for a general many-body observable $\op{O}$, the situation is more complicated,
and GGE can be proven to correctly predict  long-time averages under two additional requirements: 
({1}) $\op{O}$ is a finite sum of powers of some fermionic operators, and 
({2}) the time fluctuations of the one-body Green's functions associated to such fermionic operators are vanishing.
Whenever either of the two conditions is not realized,  $\ggeav{\op{O}}$ is not guaranteed to coincide with
$\timeav{\op{O}}$: we will indeed discuss cases of a definite disagreement between the two averages
because of a violation of condition ({2}) above, i.e., the persistence of time fluctuations of one-body Green's functions. 

The key to the story is Wick's theorem \cite{Fetter:book}, which clearly applies to the free-fermion Hamiltonians we are discussing \cite{Cazalilla2012}.
Any many-body observable $\op{O}$ can be expressed as a linear combination of powers of the original real-space fermions
$\op{c}_j$ and $\op{c}_j^\dagger$.
Since $\ket{\Psi_0}$ is a BCS-Slater determinant, we can expand $\bra{\Psi(t)} \op{O} \ket{\Psi (t)}$, using Wick's theorem, 
as a sum of products of one-body Green's functions
$G_{j_1j_2}(t) \equiv \bra{\Psi(t)} \op{c}^\dagger_{j_1} \op{c}_{j_2} \ket{\Psi(t)}$ and
$F_{j_1j_2}(t) \equiv \bra{\Psi(t)} \op{c}^\dagger_{j_1} \op{c}_{j_2}^\dagger \ket{\Psi(t)}$.
To make things more clear, let us consider, for instance, the density-density correlations 
$\op{\rho}_{j_1j_2} = \op{n}_{j_1} \op{n}_{j_2}$ with $j_1 \neq j_2$, a two-body operator whose Wick's expansion reads:
\begin{eqnarray} \label{eq:rhoexpansion}
\rho_{j_1j_2} (t) &=&  \bra{\Psi (t)} \op{c}_{j_1}^\dagger  \op{c}_{j_1} \op{c}_{j_2}^\dagger \op{c}_{j_2} \ket{\Psi (t)} \\
& =& G_{j_1 j_1}(t) G_{j_2 j_2}(t) - |G_{j_1 j_2}(t)|^2 + |F_{j_1 j_2}(t)|^2 \;. \nonumber
\end{eqnarray}
This expansion clearly involves a finite number of terms (condition ({1})). 
Now suppose (condition ({2})) that the time fluctuations of the Green's functions vanish for large $t$, hence the limits 
$\lim_{t\to \infty}G_{j_1j_2}(t)=G_{j_1j_2}(\infty)$ and $\lim_{t\to \infty}F_{j_1j_2}(t)=F_{j_1j_2}(\infty)$ exist. 
From the analysis of the previous section, it is obvious that such limits must coincide with the corresponding GGE averages: 
$G_{j_1j_2}(\infty)= \ggeav{\op{c}^\dagger_{j_1} \op{c}_{j_2}}=G_{j_1j_2}^{\rm GGE}$, and $F_{j_1j_2}(\infty)= \ggeav{\op{c}^\dagger_{j_1} \op{c}^\dagger_{j_2}}=F_{j_1j_2}^{\rm GGE}$.
It follows therefore that $\lim_{t\to \infty} \rho_{j_1j_2}(t)$ exists (i.e., its fluctuating part $\delta \rho_{j_1j_2}(t)$ vanishes for large $t$) and is given by:  
\begin{eqnarray} \label{eq:rho-time-limit}
\lim_{t\to \infty} \rho_{j_1j_2} (t) &=&  G_{j_1 j_1}^{\rm GGE} G_{j_2 j_2}^{\rm GGE} - |G_{j_1 j_2}^{\rm GGE}|^2 + |F_{j_1 j_2}^{\rm GGE}|^2 \nonumber \\
&=& \ggeav{ \op{\rho}_{j_1j_2} } \;,
\end{eqnarray}
where the final step uses the fact that Wick's theorem also applies to GGE averages of free-fermion Hamiltonians \cite{Fetter:book}.
Notice that, since long-time fluctuations of $\rho_{j_1j_2}(t)$ vanish, the infinite-time limit dominates the time average, and this implies:
\[ \timeav{\op{\rho}_{j_1j_2}} = \lim_{t\to \infty} \rho_{j_1j_2}(t) = \ggeav{ \op{\rho}_{j_1j_2} }  \;. \]
For a similar discussion when $\op{H}$ is a translationally invariant disorder-free Ising-$XX$ chain, see Ref.~\onlinecite{Cazalilla2012}.
A similar proof works quite generally for all observables $\op{O}$ provided the two stipulated conditions are satisfied.
A few important remarks are in order:
({\it i}) The existence of {\em time limits} of Green's functions, as opposed to time averages, are crucial in applying Wick's theorem, 
because it is generally false that the ``time average of a sum of products'' coincides with the ``sum of products of time averages'';
({\it ii}) For definiteness, we have chosen, above, the real-space fermions $\op{c}_j^\dagger$ to expand $\op{O}$,  
but similar arguments can be made in any one-body fermionic basis $\op{f}^\dagger_l$, for instance, a momentum 
space basis. Notice, in this respect, that $\op{O}$ might involve an infinite expansion in terms of the 
$\op{c}_j^\dagger$'s and a finite one in terms of the $\op{f}_l^\dagger$'s (condition ({1}) ). 
In this case, if the time fluctuations of 
$\mathcal{G}_{lm}(t) \equiv \bra{\Psi (t)} \op{f}^\dagger_l \op{f}_m \ket{\Psi (t)} $ and 
$\mathcal{F}_{lm}(t)  \equiv \bra{\Psi (t)} \op{f}^\dagger_l \op{f}^\dagger_m  \ket{\Psi (t)}$ vanish (condition ({2}))
one can still conclude that $\timeav{\op{O}}=\ggeav{\op{O}}$;
({\it iii}) Whenever the time fluctuations of the one-body Green's functions do not vanish for large $t$, and/or the
expansion of the operator $\op{O}$ involves an infinite number of Wick's contractions there is no guarantee
that GGE will not reproduce long-time averages: we simply cannot prove it by using Wick's theorem. 
Nevertheless, we will later discuss (see Sec.~\ref{sec:many-body-hop}) explicit cases where the persistence of one-body time fluctuations,
due to disorder and to the presence of localized eigenstates, indeed leads to a definite discrepancy between 
$\timeav{\op{O}}$ and $\ggeav{\op{O}}$.

\section{1D spinless fermions with long-range hopping: results} \label{sec:spinless}
%
Let us consider quantum quenches with a final Hamiltonian given by $\op{H}_{\mathrm{hop}}$, Eq.~\eqref{eq:fermham},
describing disordered one-dimensional spinless fermions with long-range hopping.
We will consider observables $\op{O}$ in real space and in momentum space, and hence we
will need to ascertain the time dependence of both real-space and momentum-space Green's functions. 
To unify the treatment of both cases, we will consider a general fermionic operator 
$\op{f}_n =  \sum_\mu u_{n\mu } \op{c}_\mu$ obtained by applying a unitary transformation $u$
(of matrix elements $u_{n\mu}$, with $u^\dagger u$ the identity) to the $\op{c}_\mu$'s which diagonalize $\op{H}_{\mathrm{hop}}$: 
for the original real-space fermions $\op{c}_j$, $u_{j\mu}$ is the real-space wave function of the $\mu$ eigenstate,
while for the momentum space fermions $\op{c}_k$, $u_{k \mu} = \sum_j e^{-ikj} u_{j\mu}/\sqrt{L}$. 
As discussed in Secs.~\ref{sec:GGE-one-body} and \ref{sec:GGE-many-body}, if the Green's functions
associated to the $\op{f}_n$'s have vanishing long-time fluctuations, then also the long-time 
fluctuations of $\op{O}$ disappear and $\timeav{ \op{O} } =\ggeav{ \op{O}}$. 
We will show the crucial role played by the localization of the eigenfunctions, which we will characterize through
the standard inverse participation ratio (IPR)~\cite{Aoki}. 
When quenching to a final Hamiltonian with $\alpha$ in the localized phase ($\alpha>1$), the real-space
Green's functions $G_{j_1j_2}(t)$ will be shown to have persistent time fluctuations, i.e., $\delta^2_{G_{j_1j_2}}>0$
in the thermodynamic limit $L\to \infty$, while quenching with $\alpha$ in the extended phase 
($\alpha<1$) leads to vanishing time fluctuations, $\delta^2_{G_{j_1j_2}}\to 0$.
In both cases, however, the eigenfunctions appear to be {\em extended} when analyzed in momentum space,
which results in momentum space Green's functions with vanishing time fluctuations, $\delta^2_{G_{k_1k_2}} \to 0$.
We will then explicitly discuss (see Sec.~\ref{sec:many-body-hop}) the discrepancy between time averages and GGE averages for real-space
many-body operators, such as density-density spatial correlations, when quenching to the localized phase.

\subsection{One-body Green's function fluctuations}\label{sec:G-hop}
%
As discussed in Sec.~\ref{sec:GGE-one-body}, GGE correctly predicts the infinite-time average
of any one-body operator, and hence, in particular, of the one-body Green's functions.
%
\begin{figure}[tbh]
   \begin{center}
	\includegraphics[width=0.42\textwidth]{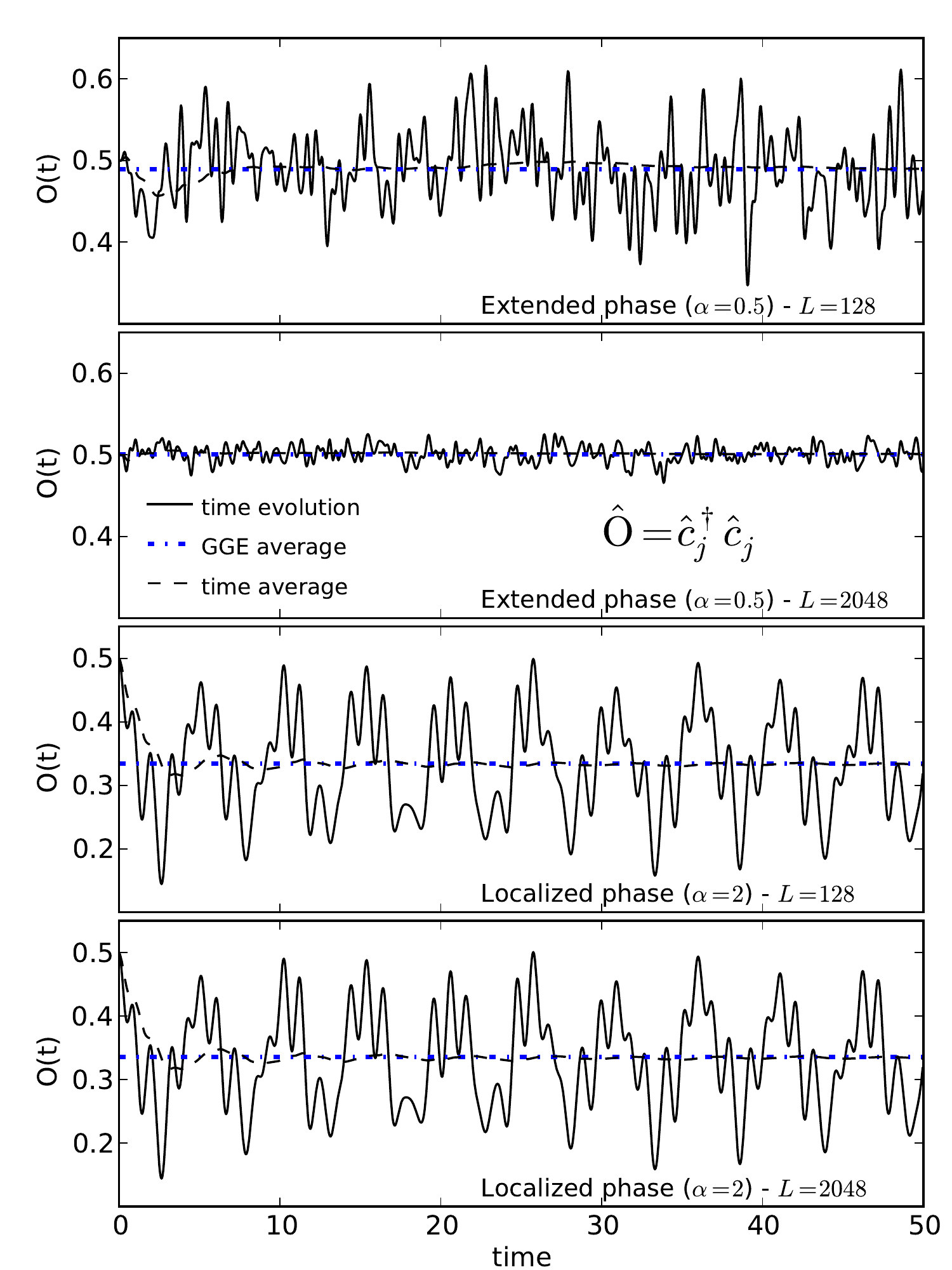}
   \end{center}
\caption{
Time evolution of $G_{jj}(t)$, with $j=L/2$ (solid line), and its running-time 
average $t^{-1} \int_0^t dt^\prime G_{jj}(t^\prime)$ (dashed line),
for two different values of $\alpha$ and two different sizes.
The horizontal dash-dotted line is the GGE average of $\op{n}_{j}$.
The data are obtained using a single realization, with $\alpha=0.5$ for the extended case, and $\alpha=2$ for the localized one
(which shows no size effect). 
Disorder realizations for the smaller $L$ chain are obtained, here and in the following, by removing sites from the
right and left edges of the larger chain.
}
\label{fig:te}
\end{figure}
%
Figure~\ref{fig:te} exemplifies this, showing the time evolution of $G_{jj}(t)$, the expectation value of the fermion density at site $j$, 
for two sizes and two values of $\alpha$, one in the extended phase ($\alpha=0.5$) and one in the
localized phase ($\alpha=2$). 
We now address in a more general way the question of the time fluctuations of the one-body Green's functions \cite{Ziraldo_PRL12}.
Since $\op{H}_{\mathrm{hop}}$ conserves the total number of fermions, the anomalous Green's functions 
$\mathcal{F}_{ml}(t)$ are always zero. 
Concerning $\mathcal{G}_{ml}(t)$, using the expansion $\op{f}_n =  \sum_\mu u_{n\mu } \op{c}_\mu$ 
we can express it as:
\begin{align} \label{eq:Gt-expanded}
\mathcal{G}_{ml}(t) & =\bra{\Psi (t) } \op{f}_m^\dagger \op{f}_l \ket{\Psi (t) } \nonumber \\
& = \sum_{\mu_1 \mu_2}  u_{m \mu_1}^\ast  u_{l \mu_2} e^{i (\epsilon_{\mu_1}-\epsilon_{\mu_2}) t} G_{\mu_1\mu_2}^0 \;,
\end{align}
where $G_{\mu_1\mu_2}^0 \equiv \bra{\Psi_0}  \op{c}_{\mu_1}^\dagger \op{c}_{\mu_2} \ket{\Psi_0}$ is the $t=0$
Green's function of the normal modes. 
Assuming the model has no single-particle energy degeneracy (i.e., if $\epsilon_{\mu_1}=\epsilon_{\mu_2}$ then 
$\mu_1=\mu_2$), only the diagonal terms with $\mu_1 = \mu_2$ will contribute to the infinite-time average of $\mathcal{G}_{ml}(t)$:
\begin{equation}
\timeav{ \op{f}_m^\dagger \op{f}_l  } 
= \ggeav{\op{f}_m^\dagger \op{f}_l}
= \sum_{\mu}  u_{m \mu}^\ast  u_{l \mu} G_{\mu\mu}^0 \;.
\end{equation}
Hence, the time fluctuations of $\mathcal{G}_{ml}(t)$ will be given by:
\begin{equation}\label{eq:deltaexpansion}
\delta \mathcal{G}_{ml} (t) = 
\sum_{\mu_1 \neq \mu_2}  u_{m \mu_1}^\ast u_{l \mu_2} e^{i (\epsilon_{\mu_1}-\epsilon_{\mu_2}) t} G_{\mu_1\mu_2}^0 \;.
\end{equation}
Now we calculate $\delta^2_{\mathcal{G}_{ml}}$ by squaring the previous expression
and taking the infinite-time average.
If we assume there is no gap degeneracy (i.e., if $\epsilon_{\mu_1} - \epsilon_{\mu_2\neq \mu_1} = \epsilon_{\mu_3}-\epsilon_{\mu_4\neq \mu_3}$ then 
$\mu_1=\mu_3$ and $\mu_2=\mu_4$) we arrive at\cite{Ziraldo_PRL12}:
\begin{equation} \label{eq:generaldelta}
\delta^2_{\mathcal{G}_{ml}} = \sum_{\mu_1 \neq \mu_2} \left|  u_{m \mu_1} \right|^2 \left| u_{l \mu_2} \right|^2  \left| G_{\mu_1\mu_2}^0 \right|^2 \;,
\end{equation}
which turns out to be a very useful and sharp tool in the analysis of the time fluctuations of the Green's functions. 
We will study it in various situations (different quenches and different choices of the fermionic operators $\op{f}_n$) to understand when and why 
$\delta^2_{\mathcal{G}_{ml}}$, which is always finite for any finite $L$ (due to revivals), goes to zero in the thermodynamic limit $L\to \infty$.
We will see, in this respect, the crucial role played by the weights $\left|  u_{m \mu_1} \right|^2 \left| u_{l \mu_2} \right|^2$ and by localization. 

\begin{figure}[tbh]
   \begin{center}
	\includegraphics[width=0.42\textwidth]{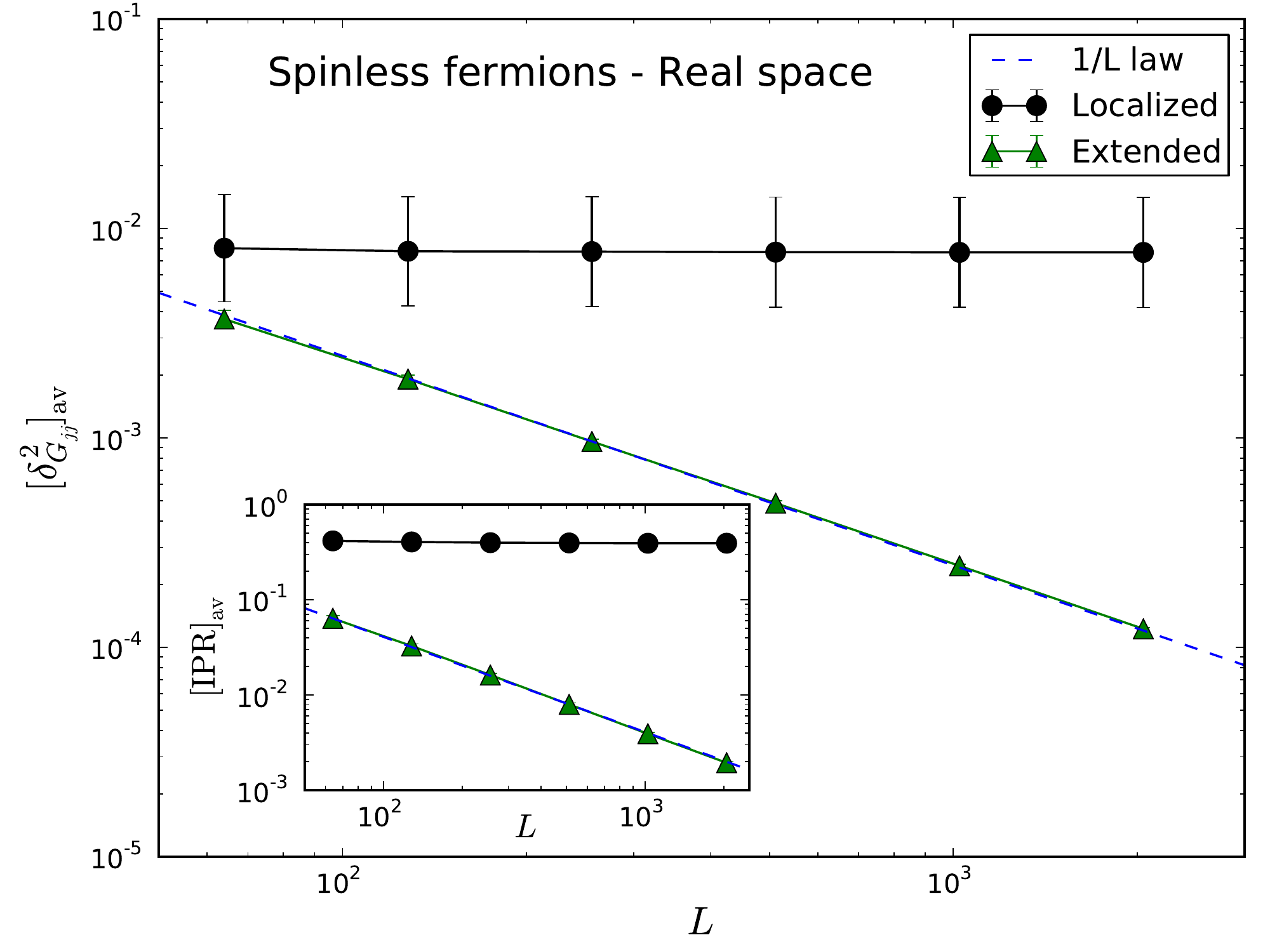}
	\includegraphics[width=0.42\textwidth]{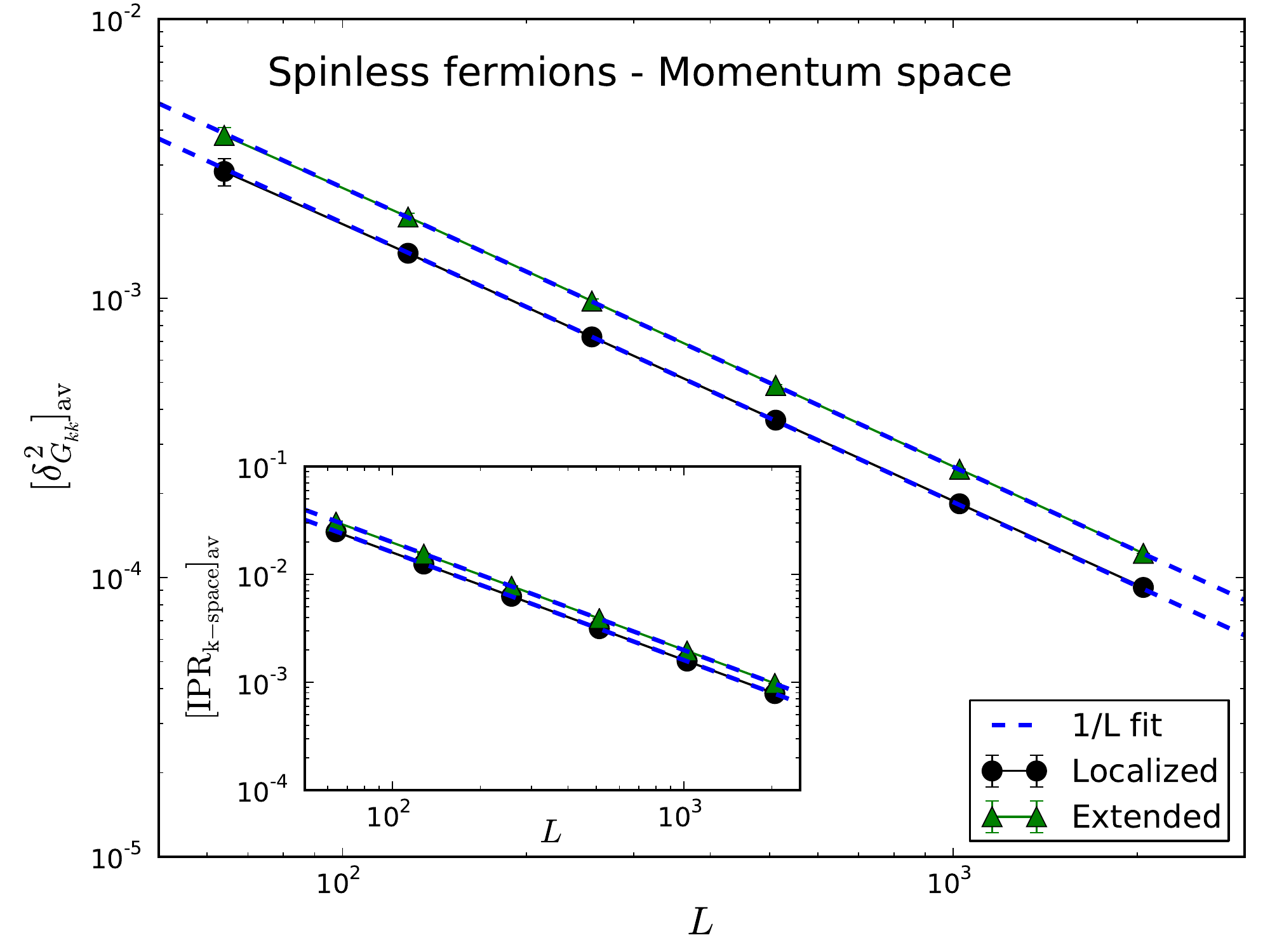}
   \end{center}
\caption{Average value of $\delta^2_{G_{jj}}$ (top) and $\delta^2_{G_{kk}}$ (bottom) 
for the disordered long-range hopping model $\op{H}_{\mathrm{hop}}$ 
as a function of the chain size $L$ for different values of $\alpha$. 
Here $G_{jj}(t)=\bra{\Psi(t)} \op{c}^\dagger_j \op{c}_j \ket{\Psi(t)}$ with $j=L/2$, and
$\mathcal{G}_{kk}(t) =\bra{\Psi (t)} \op{c}^\dagger_{k} \op{c}_{k} \ket{\Psi (t)}$ with $k=0$. 
The data are obtained starting from the ground state of a clean fermionic chain with nearest-neighbor hopping 
and quenching to $\op{H}_{\mathrm{hop}}$ with different values of $\alpha$.
The ``Localized" points are for $\alpha = 2$, while the  ``Extended" points are for $\alpha = 0.5$.
We used $50$ realizations of disorder. 
In all cases we report the usual (arithmetic) mean (the error bar, when visible, is the standard deviation), 
except for $\delta^2_{G_{jj}}$ in the localized phase, where we plot the median (the geometric mean, see Sec.~\ref{sec:models} for details).
In the inset, the IPR (see Eqs.~\eqref{eq:avIPR-realspace} and~\eqref{eq:avIPR-kspace} ) as a function of size.
Notice how for all cases (``Localized" and ``Extended" quenches) the eigenstates of the 
final Hamiltonian are ``extended'' in reciprocal space and consequently 
$\delta^2_{G_{k_1k_2}}$ goes to zero for $L\to \infty$.
For a smoother size scaling, each disorder realization of the largest $L$ generated is employed,
by removing the same amount of sites from the two edges, to generate realizations for smaller $L$.  
}
\label{fig:delta2}
\end{figure}

Let us consider first real-space Green's functions $G_{j_1j_2}(t)$, which for $j_1 = j_2 = j$, have a simple physical meaning: the expectation value at time $t$ 
of the local density at the site $j$. From Eq.~\eqref{eq:generaldelta} we get:
\begin{equation} \label{eq:delta-realspace}
\delta^2_{G_{j_1j_2}} = \sum_{\mu_1 \neq \mu_2} \left|  u_{j_1 \mu_1} \right|^2 \left| u_{j_2 \mu_2} \right|^2  \left| G_{\mu_1\mu_2}^0 \right|^2 \;,
\end{equation}
where $u_{j \mu}$ is the real-space wave functions of the eigenstate $\mu$. 
Depending on $\alpha$, the eigenfunctions $u_{j\mu}$ are either localized (for $\alpha>1$) or extended (for $\alpha<1$)~\cite{Mirlin}. 
To pin-point this, we could monitor the IPR of the $\mu$ eigenstate, ${\rm IPR}(\mu) = \sum_{ j} |u_{j \mu} |^4$. 
As it turns out, all eigenstates behave in the same way for the present model: either all localized, ${\rm IPR}(\mu) > 0$ for $L \to \infty$, 
or all extended, ${\rm IPR}(\mu) \sim 1/L$ for $L \to \infty$, without any mobility edge.
For that reason, we can just monitor the average IPR defined as:
\begin{equation}\label{eq:avIPR-realspace}
 {\rm IPR} = \frac{1}{L} \sum_{\mu} {\rm IPR}(\mu) = \frac{1}{L} \sum_{\mu} \sum_{j}  |u_{j \mu} |^4 \;,
\end{equation}
shown in the inset of Fig.~\ref{fig:delta2} (top) for $\alpha=0.5$ (``Extended'' points) and $\alpha=2$ (``Localized" points).
Correspondingly, the main panel of Fig.~\ref{fig:delta2} (top) shows the average value of $\delta^2_{G_{jj}}$ 
as a function of the chain size $L$ for both choices of $\alpha$.
Notice that when eigenstates are extended, the weights  $\left|  u_{j_1 \mu_1} \right|^2 \left| u_{j_2 \mu_2} \right|^2 \sim 1/L^2$ 
in Eq.~\eqref{eq:delta-realspace} can be essentially taken out of the sum.
But, without weights, it is a simple matter to show that:
\begin{equation}
\frac{1}{L^2} \sum_{\mu_1 \neq \mu_2} \left| G_{\mu_1\mu_2}^0\right|^2 \leq 
\frac{1}{L^2} \sum_{\mu_1  \mu_2} \left| G_{\mu_1\mu_2}^0 \right|^2 = \frac{N_\mathrm{F}}{L^2} \;,
\end{equation}
where $N_\mathrm{F}$ is the total number of fermions in the initial state. 
Hence, $\delta^2_{G_{j_1j_2}}$ is expected to go to zero as $1/L$ when quenching towards a phase with extended eigenstates ($\alpha<1$),
as indeed found numerically.
On the contrary, weights are of paramount importance when quenching to a phase with localized eigenstates ($\alpha>1$), 
because they move the important contributions to  $\delta^2_{G_{j_1j_2}}$ from the average $|G_{\mu_1\mu_2}^0  |^2$, which is of order $1/L$,  
to rare large values~\cite{Ziraldo_PRL12}, leading to a finite $\delta^2_{G_{j_1j_2}}$ which is rather insensitive to the size $L$.

Consider now the Green's functions in momentum space
$G_{k_1k_2}(t) \equiv \bra{\Psi(t)} \op{c}^\dagger_{k_1} \op{c}_{k_2} \ket{\Psi(t)} $, 
representing, for $k_1 = k_2 = k$, the expectation value at time $t$ of the momentum distribution.
Since $\op{c}^\dagger_k =  \sum_j e^{ikj} \op{c}^\dagger_j/\sqrt{L}$, 
the $G_{k_1k_2}(t)$'s are straightforwardly related to the $G_{j_1j_2}(t)$'s through
a double summation on $j_1$ and $j_2$ with oscillating phase factors $e^{i (k_1j_1-k_2j_2) }$.
However, for $L\to \infty$ these are infinite sums, and this might change the 
behavior of the time fluctuations. We now show that, even when $G_{j_1j_2}(t)$ has
persistent time fluctuations, the corresponding $G_{k_1k_2}(t)$ averages them out,
and $\delta^2_{G_{k_1k_2}}\to 0$ for $L\to \infty$.
Indeed, using Eq.~\eqref{eq:generaldelta},
\begin{equation}
\delta^2_{G_{k_1k_2}} = \sum_{\mu_1 \neq \mu_2} \left| u_{k_1 \mu_1} \right|^2
\left| u_{k_2 \mu_2} \right|^2 \left| G_{\mu_1\mu_2}^0 \right|^2 \;,
\end{equation}
where $u_{k \mu} = \sum_j e^{-ikj} u_{j\mu}/\sqrt{L}$ are the Fourier transforms of the real-space wave functions $u_{j\mu}$.
Figure~\ref{fig:delta2} (bottom) shows that $\delta^2_{G_{kk}} \to 0$ for $L\to \infty$, regardless of the value of $\alpha$.
Notice that, perhaps counter-intuitively, the average IPR in $k$-space 
\begin{equation}\label{eq:avIPR-kspace}
 {\rm IPR}_{\mathrm{k-space}} = \frac{1}{L} \sum_{\mu} \sum_{k}  |u_{k \mu} |^4 \;,
\end{equation}
always decreases as $1/L$ for both $\alpha<1$ (extended real-space wave functions) and $\alpha>1$ (localized real-space wave functions).
%
\begin{figure}[tbh]
   \begin{center}
	\includegraphics[width=0.42\textwidth]{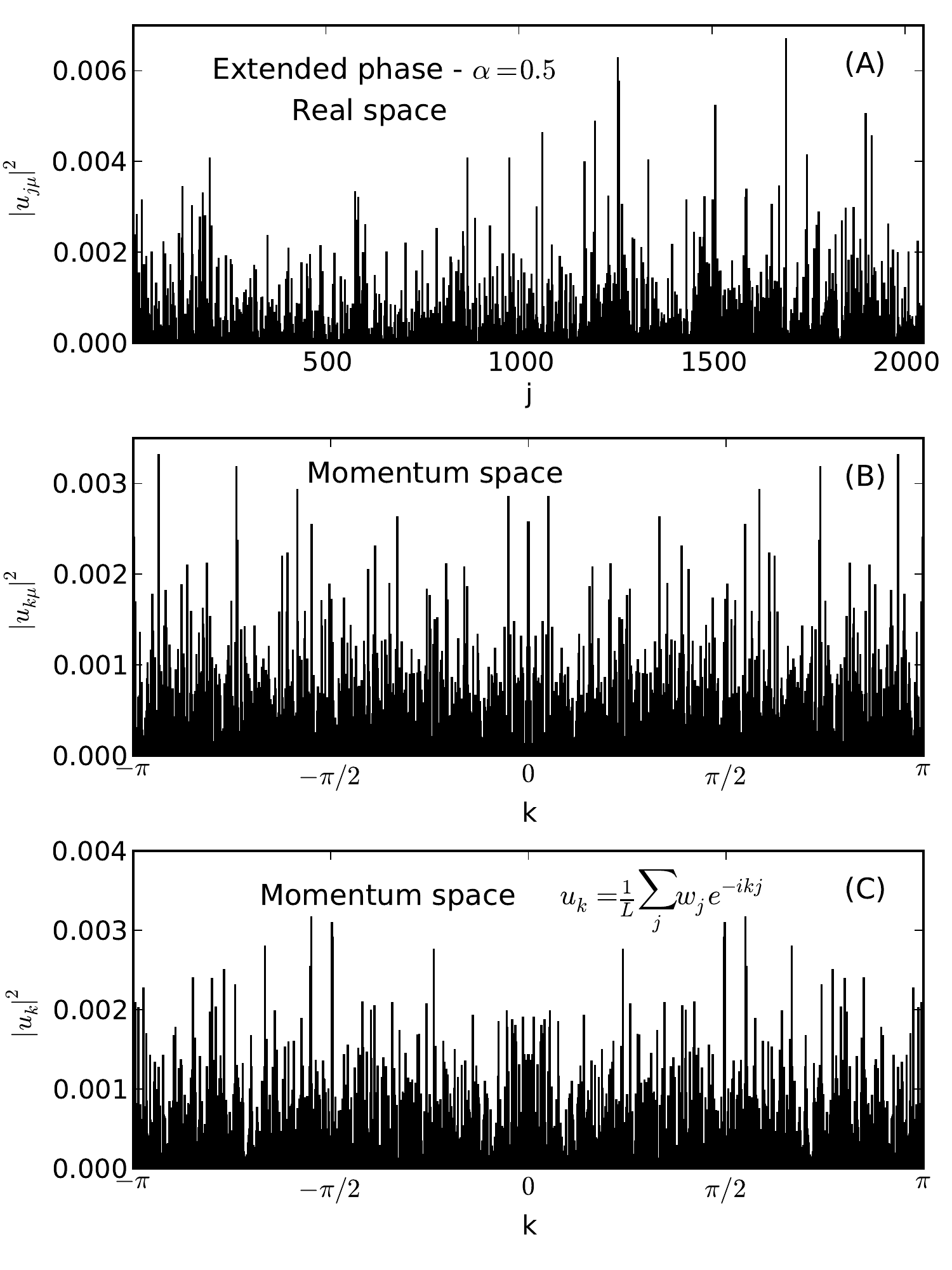}
   \end{center}
\caption{
(A) Plot of $|u_{j\mu}|^2$ versus the site index $j$, where $u_{j\mu}$ is a typical extended eigenstate of $\op{H}_{\mathrm{hop}}$ 
for $\alpha=0.5$ and $L=2048$, with energy $\epsilon_\mu$ in the middle of the band. 
(B) The corresponding momentum space $|u_{k\mu}|^2$, with $u_{k\mu} = \frac{1}{\sqrt{L}} \sum_j e^{-ikj} u_{j\mu}$, quite
clearly extended.
(C) Plot of $|u_{k}|^2$ for a toy extended wave function $u_j= w_j/\sqrt{L}$, where $w_j=\pm 1$ is a random sign.
}
\label{fig:eigenstate}
\end{figure}
%
The extended case $\alpha<1$ is particularly intriguing, because one would expect that an extended real-space wave function 
should look ``localized'' in momentum space, i.e., composed of a small number of $k$
waves. 
This expectation, quite reasonable for ordinary extended states of nondisordered systems, is
not correct, in general, in the presence of disorder, 
as quite evident from Figs.~\ref{fig:eigenstate}(A) and~\ref{fig:eigenstate}(B). 
A simple example demonstrates the crucial role played by disorder.
Consider a toy real-space extended wave function with
$u_j = w_j/\sqrt{L}$
where $w_j=\pm 1$ is a random sign on every site.
Without $w_j$, the momentum space function $u_k=\sum_j u_j e^{-ikj} / \sqrt{L}$ would be localized, with a peak at $k=0$. 
When $w_j$ is accounted for, $|u_k|^2$ becomes extremely irregular but extended over all $k$ points, see Fig.~\ref{fig:eigenstate}(C), indeed
with a strong resemblance to the actual momentum space wave function of  Fig.~\ref{fig:eigenstate}(B).
In the presence of disorder, therefore, being ``extended in real-space'' does not imply a sharply defined momentum.
Effectively, therefore, going to momentum space averages out persistent time fluctuations which are seen in real space when eigenstates are
localized, an effect akin to ``self-averaging'' of extensive quantities in disordered systems \cite{Binder_Young:review}.

\subsection{Many-body observables and failure of GGE} \label{sec:many-body-hop}

\begin{figure}[tbh]
   \begin{center}
	\includegraphics[width=0.42\textwidth]{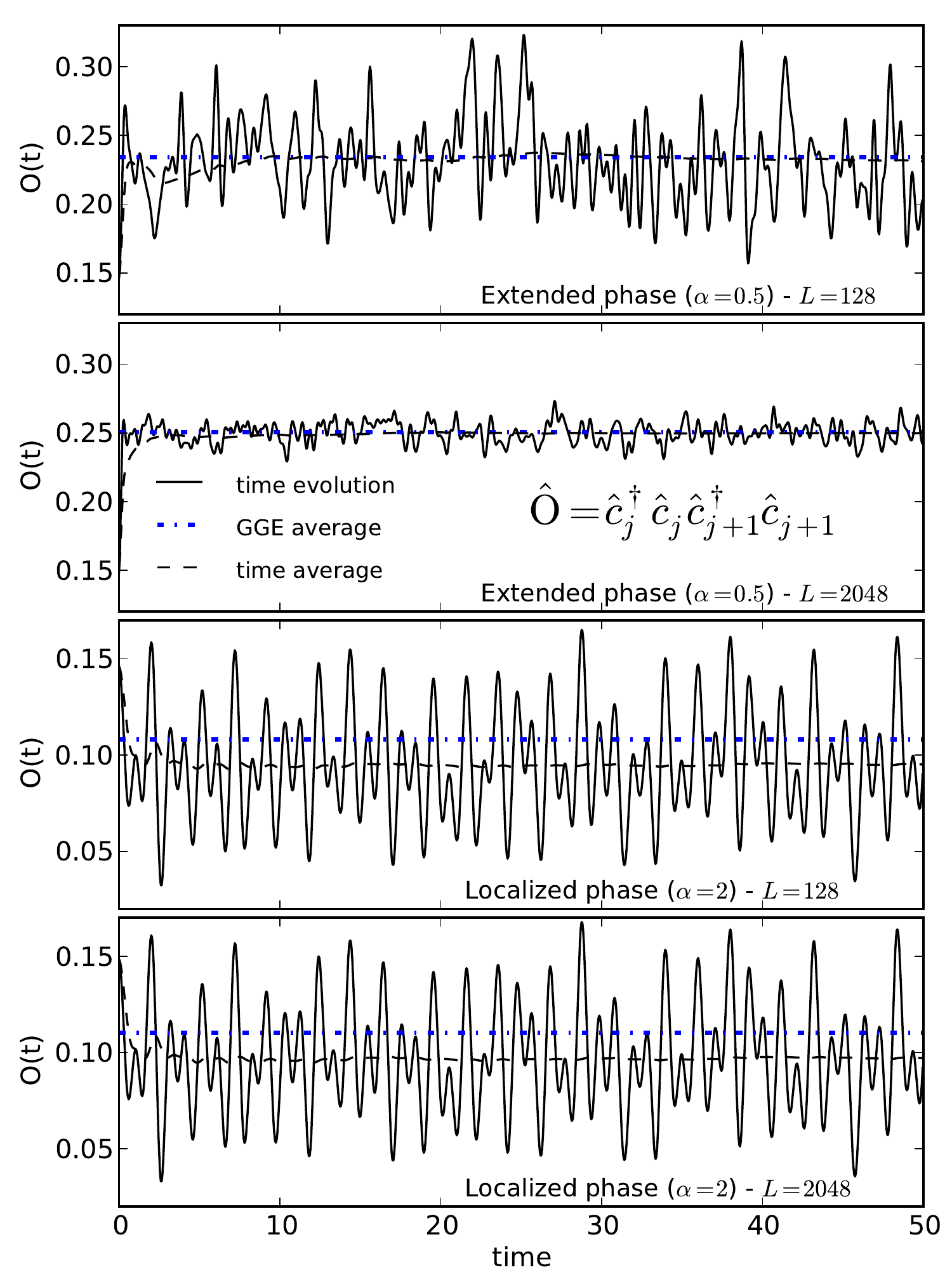}
   \end{center}
\caption{
Time evolution of $\rho_{j_1j_2}(t)$, with $j_1=L/2$ and $j_2=j_1+1$ (solid line),
and its running-time average $t^{-1} \int_0^t dt^\prime \rho_{j_1j_2}(t^\prime)$ (dashed line),
for two different values of $\alpha$ and two different sizes.
The horizontal dash-dotted line is the GGE average for $\op{\rho}_{j_1j_2}$.
The data are obtained using a single realization 
(see caption of Fig.~\ref{fig:te} for details).
}\label{fig:te-rho}
\end{figure}

From the general analysis of Secs.~\ref{sec:GGE-one-body} and \ref{sec:GGE-many-body} and the 
study of the one-body Green's functions of Sec.~\ref{sec:G-hop}, we can conclude
that many-body operators involving a finite expansion in terms of momentum space operators
$\op{c}_k$, such as correlations $\op{c}_{k_1}^\dagger \op{c}_{k_1} \op{c}_{k_2}^\dagger \op{c}_{k_2}$, 
will have time averages which coincide with GGE averages,  regardless the value of $\alpha$.
The same is true in the delocalized phase ($\alpha<1$) for many-body operators with a finite expansion 
in real space, because the time fluctuations of $G_{j_1j_2}(t)$ vanish.
When $\alpha>1$ (localized phase) the GGE ability in describing time averages of many-body operators is not guaranteed, 
because $G_{j_1j_2}(t)$ have persistent time fluctuations and Wick's theorem is of no help. 
Here we will show that, when $\alpha>1$, GGE fails in predicting the spatial density-density correlations $\op{\rho}_{j_1j_2} = \op{n}_{j_1} \op{n}_{j_2}$. 
To see this, we compare, see Eqs.~\eqref{eq:rhoexpansion} and \eqref{eq:rho-time-limit}, the time average of
$\rho_{j_1j_2}(t)$ (i.e., a time average of a sum of products of $G$'s) with the corresponding GGE average
(a sum of products of time averages).
In Fig.~\ref{fig:te-rho} we plot $\rho_{j_1j_2}(t)$, together with its running-time average (i.e., $t^{-1} \int_0^t \! dt' \rho_{j_1j_2}(t')$) and the GGE average
for two chain sizes and two values of $\alpha$. 
First we notice that, as in the case of $G_{j_1j_2}(t)$, increasing the size $L$ in the delocalized phase ($\alpha<1$) strongly 
decreases the time fluctuations, which are, on the contrary, unaffected by $L$ in the localized phase ($\alpha>1$). 
The second feature emerging from Fig.~\ref{fig:te-rho}  is that, while in the delocalized phase the time average tends to the
GGE value, there is a marked and clear discrepancy between the two in the localized phase.
%

\begin{figure}[tbh]
   \begin{center}
	\includegraphics[width=0.42\textwidth]{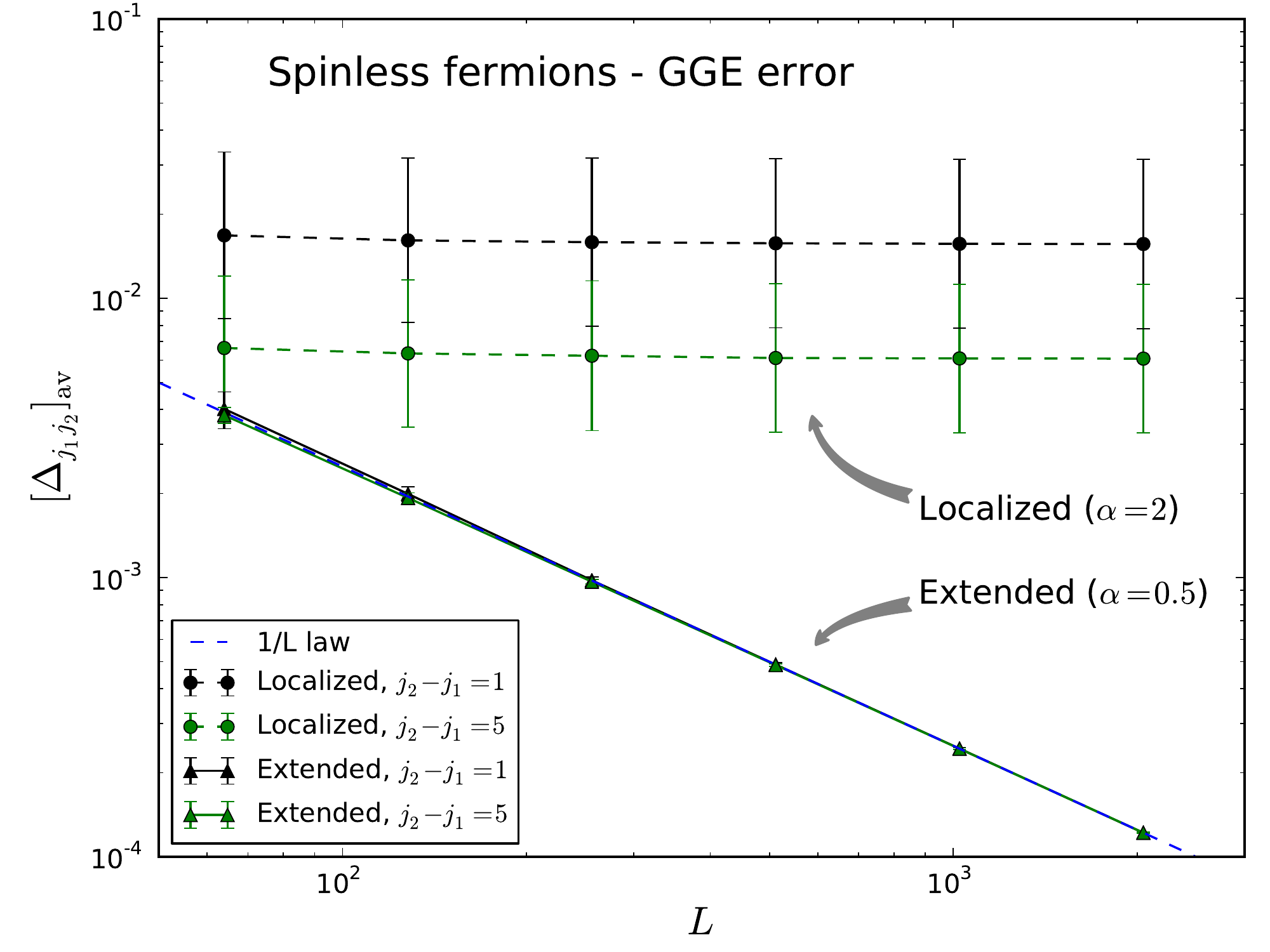}
   \end{center}
\caption{
Average value of $\Delta_{j_1j_2}= \ggeav{\op{\rho}_{j_1j_2}} - \timeav{ \op{\rho}_{j_1j_2} }$, 
the discrepancy between the GGE average and the time average, for the density-density correlation  
$\op{\rho}_{j_1j_2} = \op{n}_{j_1} \op{n}_{j_2}$,
as a function of the chain size, for with $j_1=L/2$ and two values of $j_2-j_1$.
The data are obtained using the same quenches of Fig.~\ref{fig:delta2} (see its caption for details). 
In the localized phase ($\alpha=2$) we plot the median $[\Delta_{j_1j_2}]_{\rm av}^{\rm G}= \exp ( \left[ \log \Delta_{j_1j_2} \right]_{\mathrm{av}} )$ 
because $\Delta_{j_1j_2}$ is there roughly log-normal distributed.
}
\label{fig:error}
\end{figure}

The difference between the GGE average and the time average can be explicitly computed
using the same strategy (and assumptions) of Sec.~\ref{sec:G-hop}:
\begin{widetext}
\begin{eqnarray}
\Delta_{j_1j_2} & \equiv & \ggeav{\op{\rho}_{j_1j_2}} - \timeav{ \op{\rho}_{j_1j_2} }  \nonumber \\
& = & \ggeav{ \op{c}_{j_1}^\dagger \op{c}_{j_1} } \ggeav{\op{c}_{j_2}^\dagger \op{c}_{j_2}} - 
\left| \ggeav{\op{c}_{j_1}^\dagger \op{c}_{j_2}} \right|^2 
+ \lim_{T \to \infty} \frac{1}{T} \int_0^T dt \left[  |G_{j_1j_2}(t)|^2 - G_{j_1j_1}(t) G_{j_2j_2}(t) \right]
\nonumber \\
&=&
\lim_{T \to \infty}  \frac{1}{T} \int_0^T \ud t \left[ |\delta G_{j_1j_2}(t)|^2 - \delta G_{j_1j_1}
(t) \delta G_{j_2j_2}(t) \right]
\nonumber \\
&=& \sum_{\mu_1 \mu_2} \left( | u_{j_1\mu_1} |^2 | u_{j_2\mu_2} |^2  -  u_{j_1\mu_1}^\ast u_{j_1\mu_2} u_{j_2\mu_1} u_{j_2\mu_2}^\ast \right) 
\left| G_{\mu_1\mu_2}^0 \right|^2 \;,
\end{eqnarray}
\end{widetext}
where we first used the Wick's expansions of $\ggeav{\op{\rho}_{j_1j_2}}$  and  $\bra{\Psi(t)}\op{\rho}_{j_1j_2}\ket{\Psi(t)}$,
then used the relationships $\delta G_{j_1j_2} (t) = G_{j_1j_2} (t) - \ggeav{\op{c}^\dagger_{j_1} \op{c}^\dagger_{j_2} }$ and 
$\timeav{\op{c}^\dagger_{j_1} \op{c}^\dagger_{j_2} }=\ggeav{\op{c}^\dagger_{j_1} \op{c}^\dagger_{j_2} }$ (since GGE works for one-body averages), 
and finally made use of Eq.~\eqref{eq:deltaexpansion} and of the no-gap-degeneracy assumption.
The result is closely reminiscent of Eq.~\eqref{eq:delta-realspace} for $\delta^2_{G_{j_1j_2}} $,
except that now the weights have two contributions.
Using the relation $|G_{\mu_1\mu_2}^0|^2 =|G_{\mu_2\mu_1}^0|^2$ we can finally 
reexpress $\Delta_{j_1j_2}$ as an explicitly positive quantity as follows:
\begin{widetext}
\begin{eqnarray}
\Delta_{j_1j_2} &=& \frac{1}{2} \sum_{\mu_1\mu_2}
\left[ | u_{j_1\mu_1} |^2 | u_{j_2\mu_2} |^2 + | u_{j_1\mu_2} |^2 | u_{j_2\mu_1} |^2 
-  \left( u_{j_1\mu_1}^\ast u_{j_1\mu_2} u_{j_2\mu_1} u_{j_2\mu_2}^\ast + \mathrm{c.c.}\right)  
\right]
\left| G_{\mu_1\mu_2}^0 \right|^2 \nonumber \\
&= &
\frac{1}{2} \sum_{\mu_1\mu_2}  \left|
u_{j_1\mu_1} u_{j_2\mu_2} - u_{j_1\mu_2} u_{j_2\mu_1}
  \right|^2 \left| G_{\mu_1\mu_2}^0 \right|^2  \geq 0 \;.
\end{eqnarray}
\end{widetext}
This explicitly shows that, apart from the trivial cases in which all terms of the summation are zero
(e.g., $j_1=j_2=j$, when $\op{\rho}_{jj} = \op{n}_j\op{n}_j = \op{n}_j$ becomes a one-body operator for which GGE works and $\Delta_{jj}=0$),
the GGE average overestimates the time average of $\op{\rho}_{j_1j_2}(t)$, as exemplified in Fig.~\ref{fig:te-rho} for a single realization.
The question now is whether or not $\Delta_{j_1j_2}$ 
goes to zero for $L \to \infty$.
In Fig.~\ref{fig:error} we plot the average value of $\Delta_{j_1j_2}$ 
as a function of $L$ for the same set of quenches presented before.
$[\Delta_{j_1j_2}]_{\rm av}$ goes to zero in the thermodynamic limit in the ``Extended" phase,
and this confirms our general analysis: the real-space Green's functions have vanishing 
fluctuations in the extended phase, and GGE works for many-body observables with a finite expansion.
On the contrary, in the ``Localized" phase, $[\Delta_{j_1j_2}]_{\rm av}^{\rm G}= \exp ( \left[ \log \Delta_{j_1j_2} \right]_{\mathrm{av}} )$ remains finite 
even in thermodynamic limit, and since $[\Delta_{j_1j_2}]_{\rm av} \geq [\Delta_{j_1j_2}]_{\rm av}^{\rm G}$ (by Jensen's inequality),
this ensures that also $[\Delta_{j_1j_2}]_{\rm av}$ is finite for $L \to \infty$.
Clearly,  the persistent time fluctuations of the Green's functions lead to time {\em correlations} between the $G_{j_1j_2}(t)$ appearing in the expansion 
of $\rho_{j_1j_2}(t)$, see Eq.~\eqref{eq:rhoexpansion}, which reduce the time average with respect to the corresponding ``sum of products'' of time averages.
%

\section{Ising chain in transverse field: results} \label{sec:Ising}
%
Consider now the case of an Ising chain in transverse field, described in Sec.~\ref{sec:models}.
The most important difference with respect to the case of spinless fermions with long-range hoppings is the presence of particle non-conserving
(BCS-like) terms, which mix particles and holes. 
Because of that, it is convenient to work with Nambu vectors\cite{Young1996,Young1997}, which make the algebra very similar to the previous one,
with very similar results. In particular, we will show that, due to disorder and localization of eigenstates,  the time fluctuations of 
the local magnetization $\op{\sigma}_j^z = 2 \op{c}_j^\dagger \op{c}_j - 1$ remain finite; however, considering for instance, the total magnetization $\op{m}_z = \sum_j \op{\sigma}_j^z /L$ which is extended over the whole chain, 
introduces an infinite summation which effectively leads to a self-averaging of time
fluctuations.

Consider, for simplicity, only real-space Green's functions. 
In terms of the real-space fermion operators, $\op{c}_j$ defines the Nambu vector
$\op{\Psi} \equiv \left( \op{c}_1, \dots, \op{c}_L, \op{c}_1^\dagger, \dots, \op{c}_L^\dagger \right)^\mathrm{T} $.
Similarly, in terms of the fermions $\op{\gamma}_{\mu}$ which diagonalize $\op{H}_{XY}$, we define a second Nambu vector 
$\op{\Gamma} \equiv \left( \op{\gamma}_1, \dots, \op{\gamma}_L, \op{\gamma}_1^\dagger, \dots, \op{\gamma}_L^\dagger \right)^\mathrm{T} $.
These two vectors are connected through the relation $\op{\Psi} = U \op{\Gamma}$ where 
$U$ is a $2L \times 2L$ unitary matrix which describes the Bogoliubov rotation\cite{Young1996,Young1997} performed to diagonalize $\op{H}_{XY}$.  

\begin{figure}[tbh]
   \begin{center}
	\includegraphics[width=0.42\textwidth]{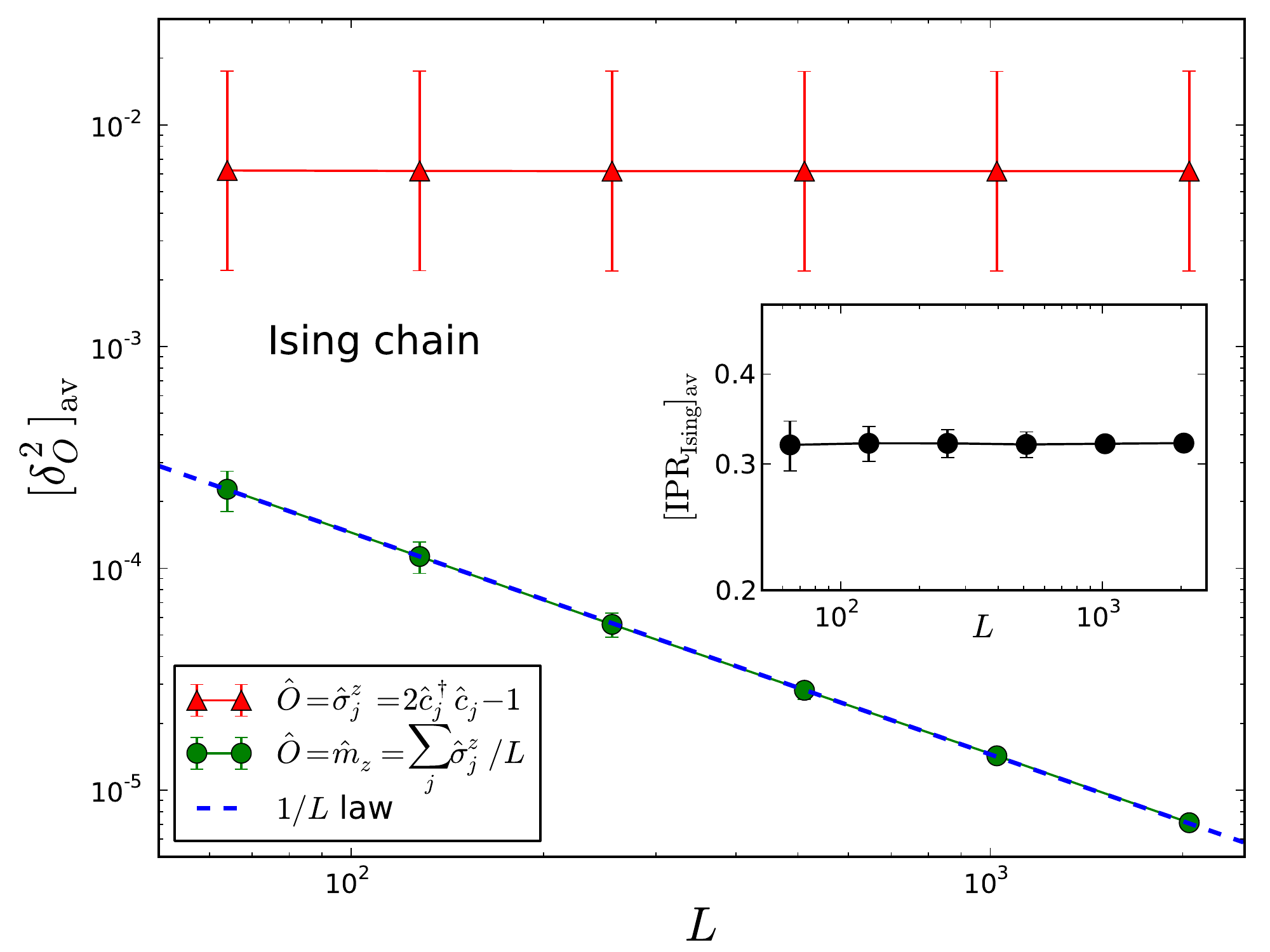}
   \end{center}
\caption{Value of $\delta^2_{O}$, with $\op{O}$ the local transverse 
magnetization $\op{\sigma}_{j}^z$ or the total magnetization $\op{m}_z$,
as a function of the chain size.
The initial state is the ground state of a clean Ising chain at the critical point  ($\epsilon = 0$, $\gamma=1$, $J_j=1$, and $h_j=1$)
and the final Hamiltonian is a disordered Ising chain at the infinite-randomness critical point  ($\epsilon = 1$, $\gamma=1$, $J_j\in [0,2]$, and $h_j\in [0,2]$).
Data obtained with $50$ disorder realizations.
For a smoother size scaling, each disorder realization of the largest $L$ generated is employed,
by removing the same amount of sites from the two edges, to generate realizations for smaller $L$. 
For $\op{O} = \op{\sigma}_{j}^z$ we plot the median 
$[\delta^2_{\sigma_j^z}]_{\rm av}^{\rm G}= \exp ( \left[ \log \delta^2_{\sigma_j^z} \right]_{\mathrm{av}} )$
because $\delta^2_{\sigma_j^z}$ is roughly log-normal distributed.}
\label{fig:ising}
\end{figure}

Notice that the index $\mu$ in the Nambu vector $\op{\Gamma}_\mu$ runs from $1$ to $2L$: for every index $\mu \in [1,L]$ 
(associated to $\op{\gamma}_{\mu}$, and with energy $\epsilon_\mu>0$) there is a corresponding index 
$\bar{\mu}=\mu+L\in [L+1,2L]$ (associated to $\op{\gamma}^{\dagger}_{\mu}$, and with energy $-\epsilon_\mu$);
similarly, for every $\mu \in [L+1,2L]$ we define $\bar{\mu}=\mu-L\in [1,L]$.
The time evolution of $\op{\Gamma}_\mu$ is simple:  
$e^{i\op{H}t} \op{\Gamma}_\mu e^{-i\op{H}t} = e^{- i \tilde{\epsilon}_{\mu} t} \op{\Gamma}_\mu$,
where the energy $\tilde{\epsilon}_\mu = \epsilon_\mu$ when $1 \leq \mu \leq L$  and
$\tilde{\epsilon}_\mu = -\epsilon_{\bar{\mu}}$ when $L+1 \leq \mu \leq 2L$.
The pairs $\op{c}^\dagger_{j_1}\op{c}_{j_2}$ and  $\op{c}_{j_1}^\dagger \op{c}_{j_2}^\dagger$ entering in the
one-body standard and anomalous Green's functions can both be obtained from the Nambu pairs 
$\op{\Psi}_{j_1}^\dagger \op{\Psi}_{j_2}$ (where now $j_1$ and $j_2$ run from $1$ to $2L$): 
$\op{\Psi}_{j_1}^\dagger \op{\Psi}_{j_2} = \sum_{\mu_1\mu_2} U^\ast_{j_1\mu_1} U_{j_2\mu_2}
\op{\Gamma}^\dagger_{\mu_1} \op{\Gamma}_{\mu_2}$,
and the associated (Nambu) Green's function closely resembles Eq.~\eqref{eq:Gt-expanded}:
\begin{equation}\label{eq:G-XY}
\mathcal{G}_{j_1j_2}(t) = \sum_{\mu_1\mu_2} 
U^\ast_{j_1\mu_1} U_{j_2\mu_2} e^{ i (\tilde{\epsilon}_{\mu_1} - \tilde{\epsilon}_{\mu_2}) t} {\mathcal G}_{\mu_1\mu_2}^0 \;,
\end{equation}
where ${\mathcal G}_{\mu_1\mu_2}^0 = \bra{\Psi_0} \op{\Gamma}^\dagger_{\mu_1} \op{\Gamma}_{\mu_2} \ket{\Psi_0}$.
Assuming, again, no energy degeneracy, the time fluctuations of  $\mathcal{G}_{j_1j_2}(t)$ read:
\begin{equation} \label{eq:fluctBCS}
\delta \mathcal{G}_{j_1j_2}(t) = \sum_{\mu_1 \neq \mu_2} U^\ast_{j_1\mu_1} U_{j_2\mu_2} e^{ i (\tilde{\epsilon}_{\mu_1} - \tilde{\epsilon}_{\mu_2}) t} 
{\mathcal G}_{\mu_1\mu_2}^0 \;.
\end{equation}
In computing the time-averaged squared fluctuations of $\mathcal{G}_{ml}(t)$ 
we have to take care of gap degeneracies due to the particle-hole symmetry of the spectrum $\tilde{\epsilon}_\mu$.
Taking due care of that, the value of $\delta^2_{\mathcal{G}_{j_1j_2}}$ turns out to be:
\begin{widetext}
\begin{align} \label{eq:bcsdelta}
\delta^2_{\mathcal{G}_{j_1j_2}} & = 
\sum_{\mu_1 \neq \mu_2} \sum_{\nu_1 \neq \nu_2} 
\overline{ e^{ i (\tilde{\epsilon}_{\mu_1} - \tilde{\epsilon}_{\mu_2} - \tilde{\epsilon}_{\nu_1} + \tilde{\epsilon}_{\nu_2}) t}}
 U^\ast_{j_1\mu_1} U_{j_2\mu_2} U_{j_1\nu_1} U_{j_2\nu_2}^\ast 
{\mathcal G}_{\mu_1\mu_2}^0 {\mathcal G}_{\nu_2\nu_1}^0
\nonumber \\
& = \sum_{\mu_1 \neq \mu_2} |U_{j_1\mu_1}|^2 |U_{j_2\mu_2}|^2 \left| {\mathcal G}_{\mu_1\mu_2}^0 \right|^2
- \sum_{ \substack{ \mu_1 \neq \mu_2 \\ \mu_1 \neq \overline{\mu_2} } } 
U_{j_1\mu_1}^\ast U_{j_2\mu_2} U_{j_1 \overline{\mu_2}} U_{j_2 \overline{\mu_1}}^\ast  
\left| {\mathcal G}_{\mu_1\mu_2}^0 \right|^2
\end{align}
\end{widetext}
where the over-line denotes the infinite-time average, and we used the relation 
$\op{\Gamma}_{\mu} = \op{\Gamma}_{\overline{\mu}}^\dagger$. 
The first term is due to the cases in which $\mu_1 = \nu_1$ and $\mu_2 = \nu_2$, similarly to what is found in
Eq.~\eqref{eq:generaldelta}. The second term originates from particle-hole symmetry 
(present even when the system is disordered)
and occurs when $\mu_1  = \overline{\nu_2}$ and $\mu_2  = \overline{\nu_1}$.
As we did for the spinless fermionic Hamiltonian, one can show that:
\begin{equation}
\sum_{\mu_1 \mu_2} |\bra{\Psi_0} \op{\Gamma}^\dagger_{\mu_1} \op{\Gamma}_{\mu_2} \ket{\Psi_0}|^2 = L \;.
\end{equation}
Moreover, using the fact that $U^\dagger U = 1$ we can repeat the same observations presented in Sec.~\ref{sec:G-hop}: 
if the eigenstates are delocalized (clean chain case) $\delta^2_{\mathcal{G}_{j_1j_2}}$ goes to zero for $L\to \infty$, 
while  $\delta^2_{\mathcal{G}_{j_1j_2}}$ remains finite when the eigenstates are localized.
For any finite disorder amplitude $\epsilon$, the Hamiltonian $\op{H}_{XY}$ has always localized states.
Indeed, as shown in the inset of Fig.~\ref{fig:ising}, the average IPR defined similarly to Eq.~\eqref{eq:avIPR-realspace},
\begin{equation}
\mathrm{IPR}_{\rm Ising} =
\frac{1}{2L} \sum_{\mu=1}^{2L} \sum_{j=1}^{2L}  |U_{j \mu} |^4 \;,
\end{equation}
is finite for $L \to \infty$.
This localization leads to persistent time fluctuations: $\delta^2_{\mathcal{G}_{j_1j_2}}>0$ in the thermodynamic limit. 
This is shown in Fig.~\ref{fig:ising} where we plot the average value of
$\delta^2_{\mathcal{\sigma}_{j}^z} = 4 \delta^2_{\mathcal{G}_{jj}} $ for a quench
from the ground state of a clean Ising chain at the critical point ($\epsilon = 0$, $\gamma=1$, $J_j=1$, and $h_j=1$) 
to a disordered Ising chain at the infinite randomness critical point ($\epsilon = 1$, $\gamma=1$, $J_j\in [0,2]$, and $h_j\in [0,2]$) \cite{Fisher_PRB95}.
%

We now show that, while each $\op{c}^\dagger_j\op{c}_j$ has nonvanishing time fluctuations, the average magnetization per site
$\op{m}_z =\sum_j \op{\sigma}^z_j / L= 2 \sum_{j=1}^L \op{\Psi}_j^\dagger \op{\Psi}_j/L - 1$ has vanishing time fluctuations for large $L$, due
to cancellations reminiscent of self-averaging in extensive observables \cite{Binder_Young:review}.
Indeed, $\delta m_z (t) = 2 \sum_{j=1}^L \delta \mathcal{G}_{jj}(t)/L$, which implies that:
\begin{equation} \nonumber
\delta m_z (t) = \sum_{\mu_1 \neq \mu_2}
\left(  \frac{2}{L} \sum_{j=1}^L U_{j\mu_1}^\ast U_{j\mu_2}  \right)
e^{i(\tilde{\epsilon}_{\mu_1}- \tilde{\epsilon}_{\mu_2})t}
{\mathcal G}_{\mu_1\mu_2}^0 \;,
\end{equation}
i.e., an expression entirely similar to Eq.~\eqref{eq:fluctBCS} for $\delta \mathcal{G}_{j_1j_2}(t)$ except for the weight
$U^\ast_{j_1\mu_1} U_{j_2\mu_2}$ which is now replaced by the 
averaged weight $w_{\mu_1 \mu_2}=2\sum_{j=1}^L  U_{j\mu_1}^\ast U_{j\mu_2}/L$.
With the same steps done to obtain Eq.~\eqref{eq:bcsdelta}, we finally get:
\begin{equation} \label{eq:delta2m}
\delta^2_{\op{m}_z } =
\sum_{\mu_1 \neq \mu_2} \left| w_{\mu_1\mu_2} \right|^2
\left| {\mathcal G}_{\mu_1\mu_2}^0 \right|^2
-  \sum_{ \substack{ \mu_1 \neq \mu_2 \\ \mu_1 \neq \overline{\mu_2} } }
w_{\mu_1\mu_2} w_{\overline{\mu_1}\overline{\mu_2}}^\ast 
\left| {\mathcal G}_{\mu_1\mu_2}^0 \right|^2 
\nonumber
\end{equation}
which is definitely different from Eq.~\eqref{eq:bcsdelta}, the site average having been performed {\em before} taking the squared time fluctuations.
In Fig.~\ref{fig:ising} we plot $\delta^2_{\op{m}_z}$, averaged over disorder realizations, as a function of the chain size $L$: 
we clearly see that, even if the eigenstates of the $\op{H}_{XY}$ are localized, the 
time fluctuations of 
$\op{m}_z$ decay, and $\delta^2_{\op{m}_z}\to 0$ for large $L$.
This behavior for $\op{m}_z$ is similar to that of the Green's functions $G_{k_1k_2}(t)$ for the disordered long-range hopping fermions
analyzed in Sec.~\ref{sec:G-hop}, where the infinite site summations lead to a cancellation of the time fluctuations of the various terms.

\section{Discussion and conclusions} \label{discussion:sec}
%
Let us discuss some of the most relevant recent papers appearing in the literature, in the light of what we have presented in our paper.
A detailed analysis of the validity of GGE averages for quantum quenches where the final Hamiltonian was integrable, disorder-free, and 
translationally invariant  (the 1D quantum Ising/$XX$ spin chains and the Luttinger model)
has been made by Cazalilla~\textit{et al.}~\cite{Cazalilla2012}, showing that, for a general class of initial states $\ket{\Psi_0}$, 
the time fluctuations of  the one-body Green's functions vanish and the GGE averages are correct, in the thermodynamic limit,  
for both local and nonlocal observables.
These results are in complete agreement with what we have shown here, since homogeneous
Hamiltonians have extended eigenstates and the time fluctuations of the one-body Green's 
functions decay for $t\to \infty$.
In our study, we have extended the analysis of Ref.~\onlinecite{Cazalilla2012} to quantum quenches with
a final Hamiltonian $\op{H}$ which is {\em disordered}: we have shown that the localization properties 
of $\op{H}$ are crucial for the relaxation of time fluctuations and, ultimately, also for the validity of the GGE. 
In particular, we showed that, for one-body operators, infinite-time averages are always 
reproduced by GGE, while for many-body operators, the localization of eigenstates of $\op{H}$ and the ensuing absence of relaxation
of one-body real-space Green's functions are, in principle, dangerous for the validity of GGE.

Quantum quenches with integrable Hamiltonians having a transition between extended and
localized states have been analyzed in a recent work by Gramsch~\textit{et al.}~\cite{Gramsch}.
They have studied the Aubry-Andr\`e model \cite{Aubry} for hard-core bosons in a one-dimensional quasiperiodic potential,
$\op{H}=\sum_{j=1}^{L-1} (\op{b}_j^\dagger \op{b}_{j+1} + \mathrm{H.c.}) + \lambda \sum_{j=1}^L \cos(2\pi \sigma j ) \op{n}_j$, where $\op{b}_j^\dagger$ ($\op{b}_j$)
creates (annihilates) a hard-core boson at site $j$, $\op{n}_j = \op{b}_j^\dagger \op{b}_{j}$, $\sigma$ is an irrational number,
and $\lambda$ is the strength of the quasiperiodic potential. 
This model can be diagonalized by mapping it, through Jordan-Wigner \cite{Lieb_AP61}, onto a noninteracting spinless fermion chain with the same potential.
The quasiperiodic on-site potential, in the absence of a true disorder, is able to induce a transition to a phase with localized eigenstates \cite{Aubry} at a finite strength of $\lambda=2$.
Reference~\onlinecite{Gramsch} considered, among others, two operators that are particularly relevant for our discussion: the local density of bosons $\op{n}_j$,
a one-body operator in terms of Jordan-Wigner fermions, and the boson momentum distribution 
$\op{m}_k = \sum_{j_1j_2} e^{ik(j_1-j_2)}\op{b}_{j_1}^\dagger \op{b}_{j_2}/L$, 
which is, on the contrary, a many-body operator when written in terms of fermions, because of a Jordan-Wigner string \cite{Lieb_AP61}.
The numerical results of Ref.~\onlinecite{Gramsch} show that when the eigenstates of the final Hamiltonian are extended ($\lambda<2$), the time fluctuations of
both $\op{n}_j$ and $\op{m}_k$ vanish, and the GGE predicts the time averages quite well, consistently with our analysis;
when the eigenstates of the final Hamiltonian are localized ($\lambda>2$) the situation is more complex: 
the time fluctuations of $\op{n}_j$ do not relax but GGE predicts well the time average (again consistently with our analysis of Sec.~\ref{sec:GGE-one-body}), 
while the time fluctuations of $\op{m}_k$ appear to vanish, but GGE seems to fail.
The failure of GGE in predicting time averages of a many-body observable like $\op{m}_k$ when persistent time fluctuations of the one-body Green's functions
are at play (localized phase, $\lambda>2$) is perfectly in line with our results (see Sec.~\ref{sec:many-body-hop}).
What is definitely beyond our analysis, but not in contradiction with it, is the fact that the time fluctuations of $\op{m}_k$ relax for large $t$: this is likely an
effect of cancellation of fluctuations due to the summation of many terms, similar to what we have found for extensive operators (see Sec.~\ref{sec:Ising}) or for momentum space Green's functions
(see Sec.~\ref{sec:G-hop}).

Another paper quite relevant for our study is that of Khatami \textit{et al.}~\cite{Khatami}, 
where they analyze quenches with a final Hamiltonian similar to our $\op{H}_{\mathrm{hop}}$, Eq.~\eqref{eq:fermham}, 
supplemented by an interaction term  $V \sum_i (\op{n}_i -1/2)(\op{n}_{i+1} -1/2)$ which definitely breaks integrability.
Two comments are in order here. 
First, as shown in Ref.~\onlinecite{Khatami}, interactions do not change the picture dramatically: numerically, a metal-insulator transition 
occurs around $\alpha \sim 1 \div 1.2$, with a quite clear metallic phase for $\alpha \lesssim 1$, and an insulating one for  $\alpha \gtrsim 1.2$. 
Second, by comparing after-quench time averages with the microcanonical
average for the momentum distribution function $\op{c}_k^\dagger \op{c}_k$ and the 
density-density structure factor $\sum_{ml}e^{ik(l-m)}\op{n}_l\op{n}_m/L$, Ref.~\onlinecite{Khatami} shows 
that quenches in the metallic phase ($\alpha \lesssim 1$) are well described by the microcanonical ensemble, 
while thermalization appears to break down when quenching to the insulating phase ($\alpha \gtrsim 1.2$).
These results are definitely in line with what we have found, and suggest that, 
independently of the integrability of the Hamiltonian, the localization properties of $\op{H}$ are crucial for the after-quench thermalization.

Finally, let us mention a technical point related to the relaxation of time fluctuations of general many-body operators 
of nonintegrable models \cite{Reimann_PRL,Reimann_PhysScr} with the technique we have used for one-body Green's functions of free-particle models.
In principle, one could compute $\delta^2_{\hat{O}}$ for a general operator $\op{O}$, starting from Eq.~\eqref{defO:eqn}.
However, to make progress, one would need to stipulate something about gap degeneracies in the {\em many-body} spectrum, i.e.,
$E_\alpha - E_\beta = E_{\alpha^\prime} - E_{\beta^\prime}$ with $\alpha\neq\alpha^\prime$ and $\beta\neq\beta^\prime$ 
(apart from the trivial case $\alpha=\beta$ and $\alpha^\prime=\beta^\prime$). 
(The assumption of the absence of gap degeneracies is often used in the literature, and dates back to the original paper of von Neumann 
\cite{vonNeumannHTheorem,vonNeumannTranslation}, who, however, carefully stipulates it to hold only within each 
microcanonical energy shell, and not for the many-body spectrum at large.)
The condition of absence of gap degeneracies \cite{Reimann_PRL} is clearly untenable for models with noninteracting quasiparticles:  
you can produce an exponentially large number of many-body states $|\alpha^\prime\rangle$ and $|\beta^\prime\rangle$ whose
spectral gap $E_{\alpha^\prime} - E_{\beta^\prime}$ coincides exactly with $E_\alpha - E_\beta$:
simply operate on $|\alpha\rangle$ and $|\beta\rangle$ by applying, in identical fashion, 
an arbitrary number of particles and/or holes, 
$|\alpha'\rangle=\op{\gamma}^\dagger_{\mu_1} \op{\gamma}^\dagger_{\mu_2} \cdots \op{\gamma}^\dagger_{\mu_n} |\alpha\rangle$, and
$|\beta'\rangle=\op{\gamma}^\dagger_{\mu_1} \op{\gamma}^\dagger_{\mu_2} \cdots \op{\gamma}^\dagger_{\mu_n} |\beta\rangle$.
Then $E_{\alpha'}-E_{\alpha}=E_{\beta'}-E_{\beta}=\epsilon_{\mu_1} + \epsilon_{\mu_2} + \cdots \epsilon_{\mu_n}$ because quasiparticles
do not interact, and therefore $E_{\alpha^\prime} - E_{\beta^\prime}=E_\alpha - E_\beta$. 
One might argue that this proliferation of exactly degenerate gaps is a peculiarity of models with noninteracting quasiparticles \cite{Reimann_PhysScr}: 
interaction effects might change the picture completely. 
This is definitely an interesting point, which deserves further studies, but certainly also a very hard one, because the combination of disorder and interactions makes the analysis highly nontrivial. 
Nevertheless, let us mention the following simple argument. 
Suppose that quasiparticles interact, but there is still an exponentially large (in the number of particles $N$)
number of states with very small spectral gap differences $\Delta$: $E_\alpha - E_\beta = E_{\alpha^\prime} - E_{\beta^\prime} + \Delta$.
Then, all these spectral gap quasi-degeneracies will appear, effectively, as true degeneracies until a time $T\sim \hbar/\Delta$ is reached, and 
that time might indeed be very large. 

In conclusion, we have analyzed after-quench relaxation and thermalization issues for ``solvable'' models with disordered fermions (or spins) in
one dimension. Several points are still open and deserve further studies. Let us just mention two of them:
What is the role of a mobility edge, separating localized from extended states, in the after-quench relaxation?
What is the role of interactions in combination with disorder in the out-of-equilibrium dynamics of closed quantum systems?

\acknowledgments
A special thanks goes to Alessandro Silva, our former collaborator in a recent publication on the same subject, for many discussions and 
a careful reading of the manuscript.
We also thank F. Becca, M. Fabrizio, J. Marino, P. Smacchia, G. Menegoz, and A. Russomanno for discussions.

Research was supported by MIUR, through PRIN-20087NX9Y7, by SNSF, through SINERGIA Project CRSII2 136287\ 1, 
by the EU-Japan Project LEMSUPER, and by the EU FP7 under grant agreement n. 280555.

\end{document}